%% file: main.tex
\documentclass[11pt]{report}

\usepackage[a4paper,top=1in,bottom=1in,left=1in,right=1in,marginparwidth=1.75cm]{geometry}
\usepackage{amsmath}
\usepackage{cite}
\usepackage{courier}
\usepackage{caption}
\usepackage{graphicx}
\usepackage[colorinlistoftodos]{todonotes}
\usepackage[colorlinks=true, allcolors=blue]{hyperref}
\usepackage[ruled,linesnumbered]{algorithm2e}
\usepackage{float}
\usepackage{setspace}
\usepackage{subfigure}
\usepackage{url}
\usepackage{tabularx}
\usepackage[utf8]{inputenc}
\usepackage{mathptmx} 
\usepackage{microtype}
\usepackage{amsmath,graphicx,url,times,booktabs, tabularx}
\usepackage{siunitx}
\usepackage{graphicx} 
\usepackage{amsmath}
\usepackage{amssymb}
\usepackage{multirow}
\makeatletter
\def\blfootnote{\xdef\@thefnmark{}\@footnotetext}
\makeatother

\begin{document}
\include{Title/front}

\include{Title/titlepage}
\include{Title/Statement_of_Originality}
\include{Title/SupervisorDeclaration}
\include{Title/AuthorshipAttribution}

\pagenumbering{roman}
\tableofcontents

\include{Front/Abstract}
\include{Front/Acknowledgement}
\include{Front/Acronyms}

\listoffigures 
\addcontentsline{toc}{chapter}{Lists of Figures}
\newpage

\listoftables 
\addcontentsline{toc}{chapter}{Lists of Tables}
\newpage

\pagenumbering{arabic}
\include{Chapter1/Chapter1}

\include{Chapter2/Chapter2}
\include{Chapter3/Chapter3}
\include{Chapter4/Chapter4}
\include{Chapter5/Chapter5}

\bibliographystyle{unsrt}
\bibliography{Ref/References}

\include{Appendix/appendix}
\end{document}

%% file: Title/front.tex
\thispagestyle{empty}
\begin{center}
\begin{figure}[H]

\includegraphics{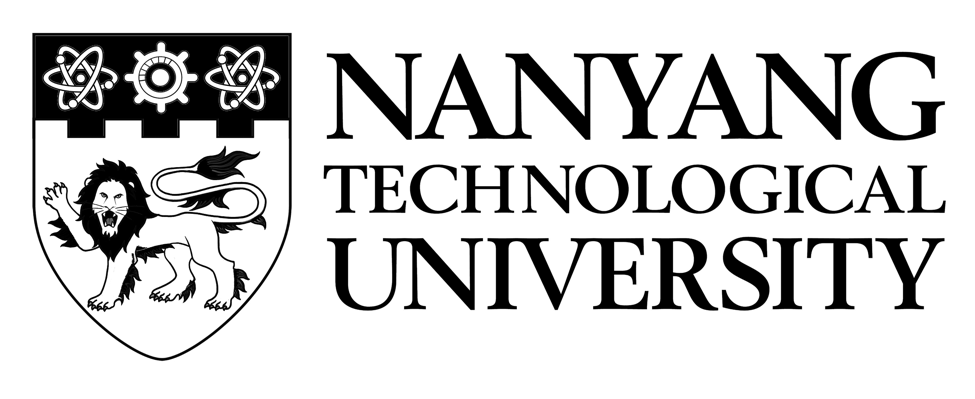}
\caption*{}
\label{fig:entropy} 
\end{figure}

\LARGE{\textbf{Continual Learning for Acoustic Event Classification}}\\[2.5in]

\large{\text{Xiao Yang}}\\[1in]

\large{\text{School of Computer Science and Engineering}}\\[0.5in]

\text{2022}
\end{center}
\newpage

%% file: Title/titlepage.tex
\begin{titlepage}

\begin{center}

\LARGE{\textbf{NANYANG TECHNOLOGICAL UNIVERSITY}}\\[2.5in]
\large{\text{MSAI Master Project MSAI/21/044}}\\[0.3in]

\LARGE{\text{END-TO-END ACOUSTIC EVENT CLASSIFICATION}}\\[2.5in]

\large{\text{Submitted by:}}\\
\large{\text{Xiao Yang}}\\
\large{\text{under the supervision of}}\\
\large{\text{Prof.Chng Eng Siong}}\\[1in]

\text{School of Computer Science and Engineering}\\[0.5in]

\text{2022}
\end{center}
\newpage

\end{titlepage}

%% file: Title/Statement_of_Originality.tex
\thispagestyle{empty}

\begin{center}

\large{\textbf{Statement of Originality}}\\[0.5in]

\begin{quote}
\large{I hereby certify that the work embodied in this report is the result of original research (and/or research survey), is free of plagiarised materials, and has not been submitted for a higher degree to any other University or Institution.}\\[1.5in]
\end{quote}

\begin{quote}
    \large{[Signature of student in this space]}\\[0.5in]
    \hrulefill\\[0.2in]
    \large{Name of Student as in Matriculation ID}\\[0.2in]
    \large{Date: [DD/MM/YYYY]}

\end{quote}


\end{center}
\newpage

%% file: Title/SupervisorDeclaration.tex
\thispagestyle{empty}

\begin{center}

\large{\textbf{Supervisor Declaration Statement}}\\[0.5in]

\begin{quote}
\large{I have reviewed the content and presentation style of this report and declare that it is free of plagiarism and of sufficient grammatical clarity to be examined. To the best of my knowledge, the research (and/or research survey) and the writing are those of the candidate except as acknowledged in the Author Attribution Statement. I confirm that the investigations were conducted in accordance with the ethics policies and integrity standards of Nanyang Technological University and that the research data are presented honestly and without prejudice.}\\[1.5in]
\end{quote}

\begin{quote}
    \large{[Signature of supervisor in this space]}\\[0.5in]
    
\begin{figure}[h]
  \centering
  \includegraphics[width=0.3\linewidth]{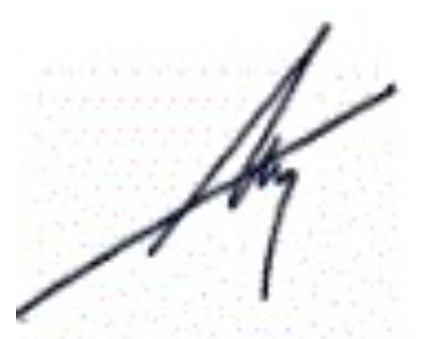}
\end{figure}

    \hrulefill\\[0.2in]
    \large{Name of supervisor as on Faculty Profile}\\[0.2in]
    \large{Date: [DD/MM/YYYY]}

\end{quote}

\end{center}

\newpage

%% file: Title/AuthorshipAttribution.tex
\thispagestyle{empty}

\begin{center}
    \large{\textbf{Authorship Attribution Statement}}\\[0.5in]

\begin{quote}
\large{Please select one of the following and strikeout the other, as appropriate:}

\begin{itemize}
  \item[](A) This report \textbf{does not} contain any material from papers published in peer-reviewed journals or from papers accepted at conferences in which I am listed as an author.
  \item[](B) This report \textbf{contains} material from the following paper(s) published in peer-reviewed journal(s) and/or accepted at conferences in which I am listed as an author.\\[0.5in]
\end{itemize}
\hrulefill
(B)
\hrulefill\\[0.5in]

\large{Please provide a list of publications (journal/conference) pertaining to the report in case you choose option (B) above. Clearly state your contribution in the work.}\\[0.4in]

\begin{enumerate}
    \item Yang Xiao*, Xubo Liu*, James King, Arshdeep Singh, Eng Siong Chng, Mark D. Plumbley. Continual Learning For On-Device Environmental Sound Classification. DCASE Workshop. (2022)\\[0.4in]
    \begin{enumerate}
        \item In this paper, my contribution is primarily in conducting the experiment and the result analysis. I have also contributed to the research in terms of implementing the methods. \\[0.4in]
        \item Xubo and I proposed the key research idea and co-prepared the manuscript drafts with James. Arshdeep contributed to the adopted BC-ResNet model. Prof. Mark D. Plumbley, Prof. Wenwu Wang, and Prof. Eng Siong Chng contributed to the revision and review of the manuscript drafts. \\[0.4in]
    \end{enumerate}

    \item Yang Xiao, Nana Hou, Eng Siong Chng. Rainbow Keywords: Efficient Incremental Learning for Online Spoken Keyword Spotting. INTERSPEECH Conference. (2022)\\[0.4in]
    \begin{enumerate}
        \item In this paper, I proposed the key research idea and implemented the proposed method as well as prepared the manuscript drafts. I have also contributed to the research in terms of conducting the experiment and the result analysis. \\[0.4in]
        \item My co-authors primarily contributed to the revision and review of the manuscript drafts. \\[0.4in]
    \end{enumerate}
\end{enumerate}

\end{quote}


\thispagestyle{empty}

\begin{quote}
    \large{[Signature of student in this space]}\\[0.5in]
    \hrulefill\\[0.2in]
    \large{Name of Student as in Matriculation ID}\\[0.2in]
    \large{Date: [DD/MM/YYYY]}

\end{quote}

\end{center}

\newpage

%% file: Front/Abstract.tex

\chapter*{Abstract}
\addcontentsline{toc}{chapter}{Abstract}

Continuously learning new classes without catastrophic forgetting is a challenging problem for on-device acoustic event classification given the restrictions on computation resources (e.g., model size,
running memory). To alleviate
such an issue, we propose two novel diversity-aware incremental learning method for Spoken Keyword Spotting and Environmental Sound Classification. Our method selects the historical data for the training by measuring the per-sample classification uncertainty. For the Spoken Keyword Spotting application, the proposed RK approach introduces a diversity-aware sampler to select a diverse set from historical and incoming keywords
by calculating classification uncertainty. As a result, the RK approach can incrementally learn new tasks without forgetting prior knowledge. Besides, the RK approach also proposes data augmentation and knowledge distillation loss function for efficient memory management on the edge device. For the Environmental Sound Classification application, we measure the uncertainty by observing how the classification probability of data fluctuates against the parallel perturbations added to the classifier embedding. In this way, the computation cost can be significantly reduced compared with adding perturbation to the raw data.\\

\par

\noindent Experimental results show that the proposed RK approach achieves 4.2\% absolute improvement in terms of average accuracy over the best baseline on \textit{Google Speech Command} dataset with less required memory. Experimental results on the \textit{DCASE 2019 Task 1} and \textit{ESC-50} dataset show that our proposed method outperforms baseline continual learning methods on classification accuracy and computational efficiency, indicating our method can efficiently and incrementally learn new classes without the catastrophic forgetting problem for on-device environmental sound classification.\\


\par

\noindent \textbf{Keywords:} Acoustic Event Classification, Environmental Sound Classification, Keyword Spotting, Continual Learning.


%% file: Front/Acknowledgement.tex

\chapter*{Acknowledgement}
\addcontentsline{toc}{chapter}{Acknowledgement}
Countless people supported my effort in this thesis. Professor Chng Eng Siong provided invaluable feedback on my analysis and framing, at times responding to emails late at night and early in the morning.  I very much appreciate the weekly meetings we had together with our team. As I was a rookie at SpeechLab@NTU, he always provide me with the correct direction for the publications. Thank you to my supervisor, Prof. Chng, for your patience, guidance, and support. I am extremely grateful that you took me on as a student and continued to have faith in me over the year.\\

\noindent Several other people gave helpful advice as I wrote, including Ng Dian Wen and so on. Dian Wen is an excellent collaborator and patient brother to me. His guidance and ideas have helped me immensely this year with the project. Besides, Chen Chen, Yuchen, and other friends in our lab also provided help and high-quality discussions with me. I would miss these days so much. I wish their research roads were better and better.\\

\noindent I especially want to thank Dr. Hou Nana. I have benefited greatly from your wealth of knowledge and meticulous editing. Without your guidance, I could never learn how to do research and published my first conference paper.\\

\noindent Thank you to my parents, for always being there for me and for telling me that I am awesome even when I didn't feel that way. Dad, thank you for all of your love and for always reminding me of the end goal. \\

\noindent Someone said: 'The giant looks in the mirror and sees nothing.' Now I will step to the next station of my life. Finally, I want to thank myself for my day and night efforts. \\

%% file: Front/Acronyms.tex

\chapter*{Acronyms}
\addcontentsline{toc}{chapter}{Acronyms}

\noindent \textbf{CNN}  Convolutional Neural Network \\

\noindent \textbf{CL} Continual Learning \\

\noindent \textbf{RK} proposed Rainbow Keyword approach \\

\noindent \textbf{RCL}  Replay-based CL \\

\noindent \textbf{MUA} Memory Update Algorithm \\

\noindent \textbf{KWS} KeyWord Spotting \\

\noindent \textbf{ASR} Automatic Speech Recognition \\
 
\noindent \textbf{FFNN} Fully-connected Feedforward Neural Network \\

\noindent \textbf{ReLU} Rectified Linear Unit \\

\noindent \textbf{MFCC} Mel-Frequency Cepstrum Coefficients \\

\noindent \textbf{DS-CNN} Depthwise separable CNN \\

\noindent \textbf{RNN} Recurrent Neural Network \\

\noindent \textbf{BiLSTM} Bidirectional LSTMs \\

\noindent \textbf{ViT}  Vision Transformer \\

\noindent \textbf{SSL} Self-Supervised representation Learning \\ 

\noindent \textbf{MISP} Multi-model Information based Speech Processing \\

\noindent \textbf{DCASE} Detection and Classification of Acoustic Scenes and Events \\

\noindent \textbf{CRNN} Convolutional Recurrent Neural Network \\

\noindent \textbf{ESC} Environmental Sound Classification \\

\noindent \textbf{GMM} Gaussian Mixture Model \\ 

\noindent \textbf{AST} Audio Spectrogram Transformer \\

\noindent \textbf{SOTA} State-Of-The-Art \\

\noindent \textbf{SSAST}  Self-Supervised AST \\

\noindent \textbf{MSPM} Masked Spectrogram Patch Modeling \\

\noindent \textbf{class-IL/CIL} Class-Incremental Learning \\

\noindent \textbf{task-IL} task-Incremental Learning \\

\noindent \textbf{MC} Monte-Carlo \\

\noindent \textbf{KD} Knowledge Distillation \\
 
\noindent \textbf{GSC} Google Speech Command dataset \\
 
\noindent \textbf{ACC} Average Accuracy \\

\noindent \textbf{BWT} Backward Transfer \\


%% file: Chapter1/Chapter1.tex

\chapter{Introduction}
\section{Background}
Audio classification refers to a series of tasks that assign labels to an audio clip \cite{liu2022learning}. There are many applications of audio classification, such as acoustic scene classification \cite{liu2022simple}, sound event detection \cite{mesaros2021sound} and keywords spotting \cite{lopez2021deep}. Audio classification is an important research topic in the field of signal processing and machine learning. Audio classification plays a key role in many real-world applications including acoustic monitoring \cite{radhakrishnan2005audio}, healthcare \cite{peng2009healthcare} and multimedia indexing \cite{kiranyaz2006generic}. \\

\noindent Neural network methods such as convolutional neural networks (CNNs) have been used for audio classification and achieved state-of-the-art performance \cite{xu2018large}. In many real-world scenarios, audio classification models need to be deployed on resource-constrained platforms such as mobile devices \cite{xiao2022continual}. Therefore, current deep-learning-based audio classification systems are usually trained with limited classes in the compact model for lower computation and smaller footprint \cite{huang2022progressive}. \\

\noindent Therefore, the performance of the model trained by the source-domain data may degrade significantly when confronted with unseen classes of the target-domain at run-time. When model developers want to expand the categories of audio to be classified, one way to do this is to fine-tune the model with new classes of data. However, this method may discard previously learned knowledge during the fine-tuning process: this is also known as the catastrophic forgetting problem \cite{mccloskey1989catastrophic}. Another possible solution is to re-train classification models with a mixture of historical and new data. However, this method is resource- and time-consuming in real-world on-device scenarios. 
As the solution based on re-training is computationally expensive, it is important to design efficient and effective methods to adapt the trained on-device audio classification model to new sound classes. \\

\noindent Continual learning (CL) \cite{awasthi2019continual} aims to continuously learn new knowledge over time while retaining and reusing previously learned knowledge. Recently, CL methods have shown promising results outperforming fine-tuning methods in deep learning tasks such as image classification \cite{mai2022online}, robotics \cite{lesort2020continual} and natural language processing \cite{biesialska2020continual}. Also, some researchers explore the approach of the continual learning for audio processing, such as \cite{huang2022progressive} and \cite{xiao2022rainbow}. However, CL in on-device applications, such as on-device audio classification, has received less attention in the literature, which is the focus in this thesis. The on-device scenarios are often associated with restrictions in storage and memory space \cite{singh2022passive}, which can pose challenges to CL which relied on external memory to restore historical data. As a result, the audio classification models that can be operated on the device may be limited in their capacities, thus prone to forgetting old knowledge when continuously learning new sound classes. \\


\section{Motivation}
Continuously learning new classes without catastrophic forgetting is a challenging problem for on-device audio classification given the restrictions on computation resources (e.g., model size, running memory). The objective of this thesis is to find the approach to solve such problem. We will investigate the state-of-the-art CL methods for classification, and then propose the most suitable approach for the different audio classification applications.\\

\section{Major contribution of the Dissertation}

In this thesis, two continual learning approaches are proposed to solve the forgetting problem for two different on-device audio classification tasks.\\

\noindent\subsection{Continual learning for keyword spotting} We proposes a novel diversity-aware continual learning approach named Rainbow Keywords (RK) to address the issues mentioned above, requiring no task-ID information with fewer parameters. Specifically, the proposed RK approach introduces a diversity-aware sampler to select few but diverse examples from historical and incoming keywords by calculating classification uncertainty. As a result, the model will not forget the prior knowledge when learning new keywords even utilizing limited historical examples. Furthermore, we utilize a mixed-labeled data augmentation to additionally improve the diversity of selected examples for higher performances. Besides, we propose a knowledge distillation loss function to guarantee that the prior knowledge could remain from the limited selected examples. We conduct our experiments on \textit{Google Speech Command} dataset following the setup of prior work \cite{mai2022online,prabhu2020gdumb}. Experimental results show that the proposed RK approach achieves 4.2\% absolute improvement in terms of Average Accuracy over the best baseline with less required memory. The scripts are available on GitHub \footnote{\url{https://github.com/swagshaw/Rainbow-Keywords}}.\\

\noindent\subsection{Continual learning for environmental sound classification} We investigate the replay-based CL (RCL) methods for on-device environmental sound classification. We first study the performance of existing memory update algorithm (MUA) methods such as \textit{Reservoir} \cite{vitter1985random}, \textit{Prototype} \cite{rebuffi2017icarl} and \textit{Uncertainty} \cite{bang2021rainbow} (as described in Section \ref{sec:mua}) on RCL for on-device environmental sound classification. \\

\noindent We empirically demonstrate that \textit{Uncertainty} \cite{bang2021rainbow} method performs best in our scenario. Furthermore, we propose \textit{Uncertainty++}, a simple yet efficient MUA method based on \textit{Uncertainty} method. Different to the \textit{Uncertainty} method, our proposed \textit{Uncertainty++} introduces the perturbations to the embedding layer of the classifier. As a result, the computation cost (e.g., running memory and time) can be significantly reduced when measuring the data uncertainty. We evaluate the performance of our method on the DCASE 2019 Task1 \cite{tau2019} and the ESC-50 \cite{piczak2015esc} datasets with on-device model BC-ResNet-Mod ($\sim$86k parameters) \cite{kim2021broadcasted,Kim2021b}. Experimental results show that \textit{uncertainty++} outperforms the existing MUA methods on classification accuracy, indicating its potential in real-world on-device audio applications. Our proposed method is model-independent and simple to apply. Our code is made available at the GitHub\footnote{\url{https://github.com/swagshaw/ASC-CL}}.

\section{Organisation of the Dissertation}

This thesis is divided into five chapters and an overview of each chapter is as follows:

\begin{itemize}
    \item Chapter 2 provides a through review of related works in two typical application of audio classification field which span from traditional approach to current deep learning approach. The chapter also explores various popular benchmarks employed for training and testing. Then chapter 2 also gives an overview of the continual learning methods. 
    \item Chapter 3 gives an overview of the proposed Rainbow Keywords method for keyword spotting. Then it provides details of the experiments based on \textit{Google Speech Commands} dataset.
    \item Chapter 4 gives an overview of the proposed Uncertainty++ method for environmental sound classification. Then we evaluate the performance of our method on the DCASE 2019 Task1 \cite{tau2019} and the ESC-50 \cite{piczak2015esc} datasets with on-device model BC-ResNet-Mod ($\sim$86k parameters) \cite{kim2021broadcasted,Kim2021b}.
    \item Chapter 5 concludes the thesis and summarizes the future work.
\end{itemize}

\newpage

%% file: Chapter2/Chapter2.tex

\chapter{Literature Review}
Audio classification is one of the prominent fields of Audio Processing. It has been widely used on multiple real world applications such as voice assistants and so on. Humans can hear sounds in to the frequency range of 20 Hz to 20 kHz based on the pressure applied on the eardrum \cite{darji2017audio}. Through the origin, we briefly split the audio classification tasks into two categories: speech and natural sound. And in this thesis, we propose the continual learning methods for the typical tasks in two categories. They are keyword spotting and environmental sound classification. \\

\noindent Hence, to provide a comprehensive literature review of research, Section \ref{sec:kws} briefly described modern deep learning based approaches to keyword spotting. The following Section \ref{sec:esc} reviewed various approaches to environmental sound classification. Then, Section \ref{sec:cl} explored the continual learning methods and their evolution.
\section{Keyword Spotting}
\label{sec:kws}
\subsection{Definition and Background}
Spoken keyword spotting (KWS) \cite{lopez2021deep} aims to identify the specific keywords in the audio input. It serves as a primary module in many real-world applications, such as Apple Siri and Google Home, which are widely utilized on the edge device. A distinguishing feature of voice assistants is that in order to use them, they must first be activated via a verbal wake word or keyword. This eliminates the need to run Automatic Speech Recognition (ASR), which is much more computationally expensive.  In particular, keyword spotting (KWS) can be defined as the task of identifying keywords in audio streams containing speech. Furthermore, in addition to voice assistant activation, KWS has many applications such as voice data mining, audio indexing, and phone routing. \cite{zhuang2016unrestricted}. Over the years, various technologies have been explored for KWS.\\
\subsection{Keyword Spotting Approach}

\begin{figure*}[htbp]
  \centering
  \includegraphics[width=\linewidth]{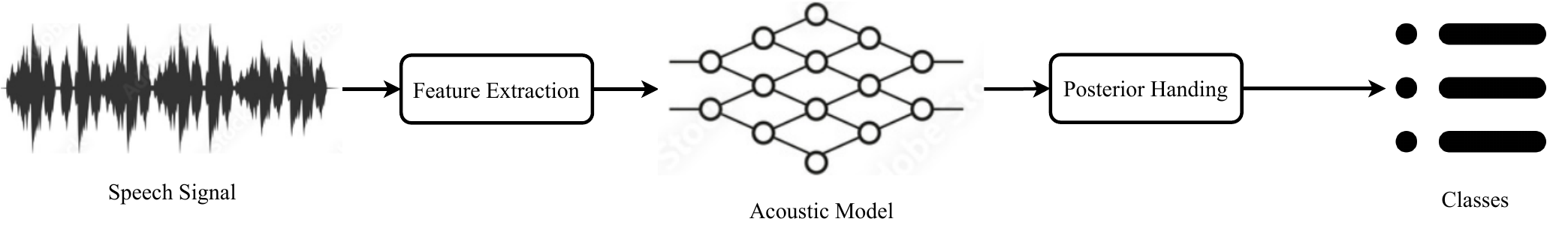}
  \caption{\textit{Overview of a keyword spotting system}}
  \label{fig:kws}
\end{figure*}

\noindent Figure \ref{fig:kws} shows the general flow of a modern deep spoken keyword spotting system. This system consists of his three main modules: It tries different keywords and stuffed (non-keyword) classes from the audio features, and 3) a post-handler processes the time-series posterior to determine the possible keywords present in the input signal.\\

\noindent This section describes the acoustic model, which is the core of the Deep Spoken KWS system. A natural tendency is to design more accurate models while reducing computational complexity.

\subsubsection{Fully-connected neural network:}
In 2014, deep spoken keyword spotting began to employ acoustic modeling based on the most popular type of neural architecture at the time. Fully Connected Feedforward Neural Network (FFNN) \cite{chen2014small}. A simple stack of three fully connected hidden layers, each with 128 neurons and rectified linear unit (ReLU) activations, followed by a softmax output layer, has fewer parameters and (at the time) states and was significantly better. the-art Keyword/fill HMM system in clean and noisy acoustic conditions. However, the use of fully connected FFNNs was quickly relegated to a secondary level due to the coherent goal of designing more accurate, robust, and less computationally intensive acoustic models. \\

\noindent Closely related and computationally cheaper alternatives
to fully-connected FFNNs are single value decomposition
filter (SVDF) \cite{alvarez2019end,liu2022learning,nakkiran2015compressing} and spiking neural networks \cite{pedroni2018small}. A closely related and less computationally expensive alternative
A fully connected FFNN has a single-value decomposition
Filters (SVDF) and spiking neural networks. Proposed in \cite{nakkiran2015compressing}, approximating the fully connected layer with a low-rank approximation, SVDF reduces the size of his FFNN acoustic model for the first deep KWS system \cite{chen2014small} by 75\% without any performance penalty. can be reduced. A spiking neural network (SNN) processes information in an event-driven manner. If such information is sparse in KWS, it significantly reduces the computational load \cite{pedroni2018small}.

\subsubsection{Convolutional Neural Network:}
\cite{sainath2015convolutional} is the signal transferred from fully connected FFNN to CNN in 2015. appreciated
CNN using local audio time-frequency correlation
For deep KWS acoustic modeling with fewer parameters, it can perform better than fully connected FFNN. one of the attractions
The characteristics of CNN are
By tuning various hyperparameters such as filters, the model can be easily constrained to meet computational constraints.
stride, and kernel and pool sizes. Also, this
It can be done without sacrificing too much performance. \\


\noindent Tang and Lin \cite{tang2018deep} are the original authors of Exploring Deep KWS Deep Residual Learning. They also integrated extended convolutions to increase the network's receptive field and capture longer time-frequency patterns without increasing the number of parameters \cite{ibrahim2019keyword,xu2020depthwise,coucke2019efficient}. Thus, Tang and Lin significantly outperform standard CNN \cite{sainath2015convolutional} in terms of the performance of KWS with fewer parameters, establishing a new state-of-the-art in 2018.\\

\begin{figure*}[htbp]
  \centering
  \includegraphics[width=0.618\linewidth]{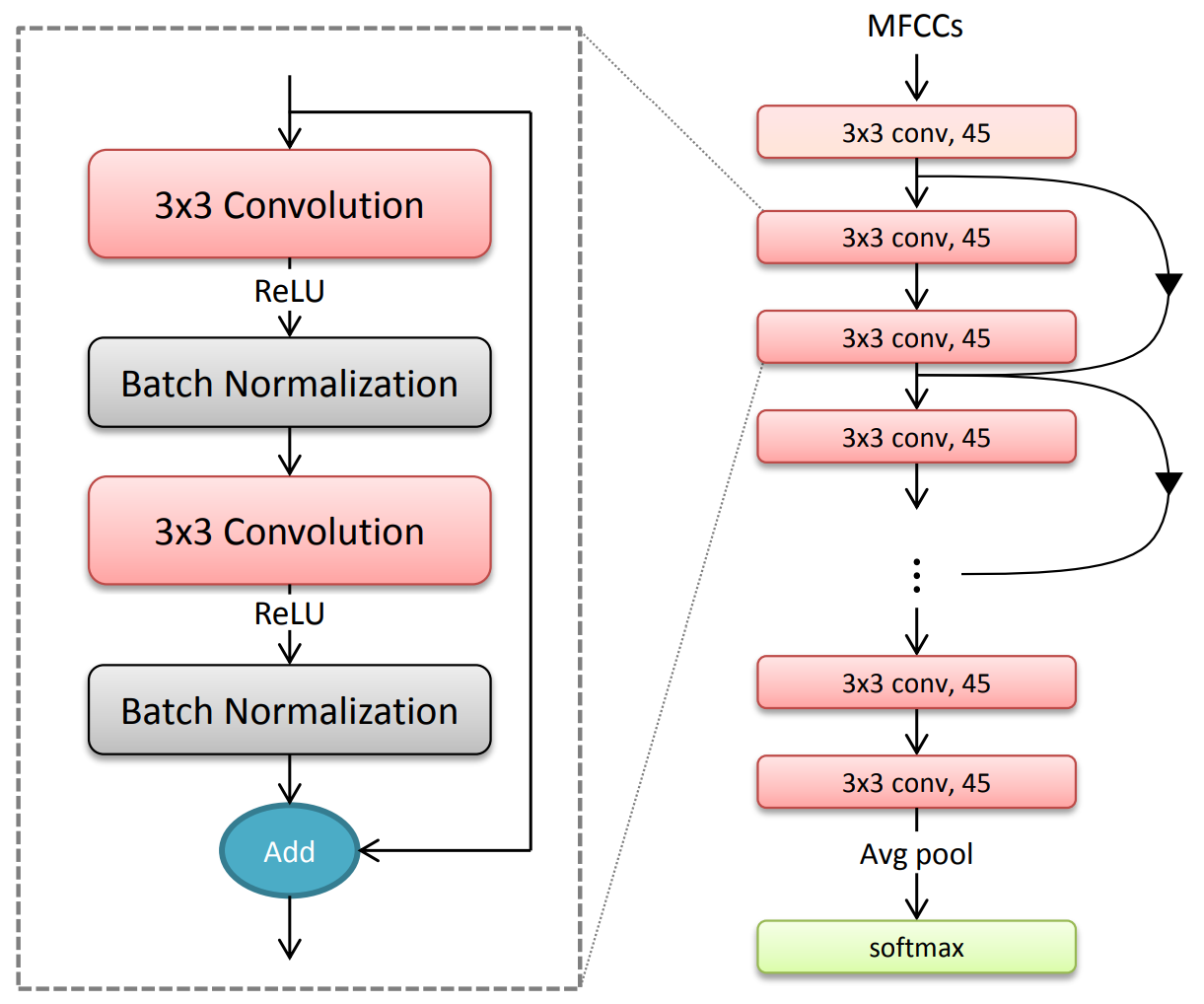}
  \caption{\textit{Full architecture, with a magnified residual block from \cite{tang2018deep}}}
  \label{fig:res15}
\end{figure*}

\noindent This success is a big motivation for further work later
The use of deep residual learning is being considered. Choi et al. \cite{choi2019temporal} characterizes the Mel-Frequency Cepstrum Coefficients (MFCC)
Input channel for the deep residual learning framework (TC-ResNet). This approach helps overcome the challenge of capturing both high- and low-frequency features through networks that are not very deep, which is also largely achieved by 2D dilated convolution, which increases the receptivity of the network. I think we can. Proposed
Compared to his 2D convolution with the same number of parameters, temporal convolution significantly reduces the computational load. As a result, TC-ResNet matches Tang and Lin's \cite{tang2018deep} KWS performance and significantly reduces latency on his mobile device \cite{choi2019temporal} and his floating point operations per second . TC-ResNet exhibits one of the lowest latencies and model sizes, outperforming KWS, standard CNNs, convolutional recurrent neural networks (CRNNs), and RNNs with attention mechanisms (see also next section) outperforms competing acoustic models based on (see also next section). Therefore, we use TC-ResNet-8 \cite{choi2019temporal} as a testbed to evaluate the proposed rainbow keyword method.\\

\noindent Another appealing way \cite{mittermaier2020small} reduces computation.
The size of a standard CNN is determined by the depthwise separable convolution. They decompose the standard convolution into depthwise and pointwise (1×1) convolutions and combine the outputs of the depthwise convolution to generate a new feature map \cite{howard2017mobilenets}. Depthwise Separable CNN (DS-CNN) is an excellent choice for implementing acoustic models with good performance in embedded systems. \\

\noindent From the review \cite{lopez2021deep}, they summarize that a modern CNN-based acoustic model should ideally encompass the following three aspects :
\begin{itemize}
    \item A mechanism to exploit long time-frequency dependencies like, e.g., the use of temporal convolutions \cite{choi2019temporal} or dilated convolutions.
    \item  Depthwise separable convolutions \cite{howard2017mobilenets} to substantially reduce both the memory footprint and computation of the model without sacrificing the performance.
    \item Residual connections to fast and effectively train
deeper models providing enhanced KWS performance.
\end{itemize}

\subsubsection{Recurrent neural network:}
Speech is a time series with strong time dependence. Therefore, it was natural to use recurrent neural networks (RNNs) for acoustic modeling. If latency is not a hard constraint, a bidirectional LSTM (BiLSTM) \cite{kumar2018convolutional,sundar2015keyword} can be used to capture causal and anti-causal relationships and improve KWS performance. Or check out KWS' bi-directional GRU at \cite{rybakov2020streaming}. When modeling doesn't take long, due to time dependencies, as in the case of KWS, the GRU is better than LSTM as it requires less memory. Train faster with similar or comparable performance Better \cite{arik2017convolutional}. \\

\noindent CNNs can have difficulty modeling long-term dependencies. To overcome this, we can combine them with RNNs to build so-called CRNNs. Therefore, CRNN offers the best of both worlds. First, a convolutional layer models the local spectral and temporal dependencies of the speech, then a recurrent layer models the long-term temporal dependencies of the speech signal by modeling follow. Some studies have investigated acoustic modeling with CRNN in deep spoken KWS using unidirectional or bidirectional LSTMs or GRUs \cite{rybakov2020streaming,arik2017convolutional,kumar2018convolutional,sundar2015keyword}.

\subsubsection{Transformer:}

\begin{figure*}[htbp]
  \centering
  \includegraphics[width=0.618\linewidth]{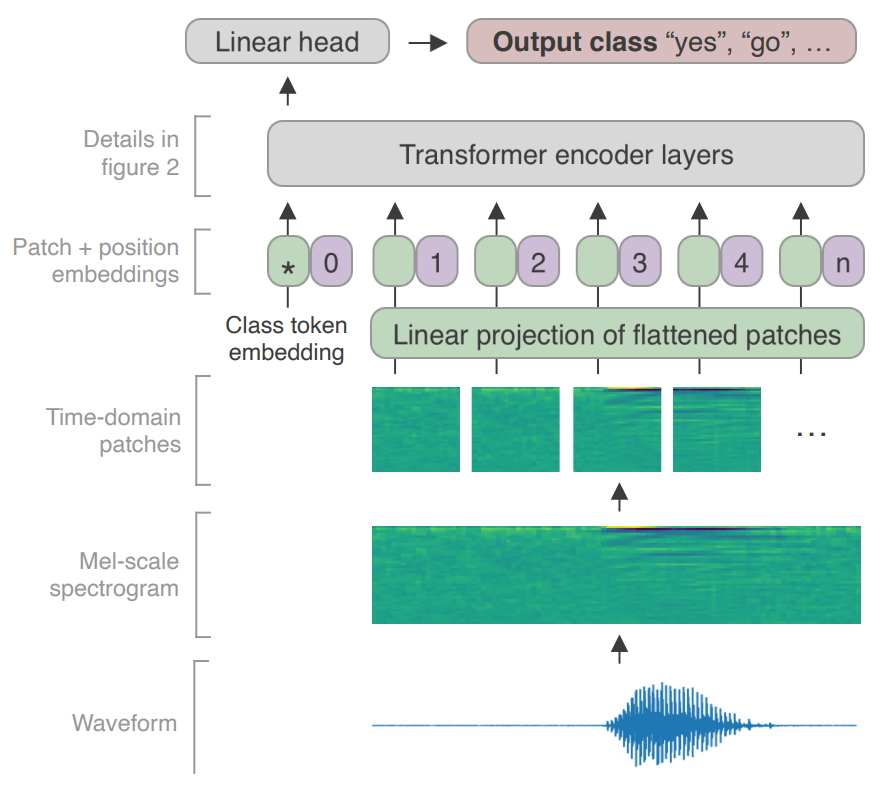}
  \caption{\textit{The Keyword Transformer architecture from \cite{berg2021keyword}. Audio is preprocessed into a mel-scale spectrogram, which is partitioned into non-overlapping patches in the time domain. Together with a learned class token, these form the input tokens for a multilayer Transformer encoder. As with ViT \cite{dosovitskiy2020image}, a learned position embedding is added to each token. The output of the class token is passed through a linear head and used to make the final class prediction.}}
  \label{fig:kwt}
\end{figure*}

 Transformer architecture have recently produced state of the art results in a variety of domains including protein sequences \cite{madani2020progen}, text \cite{brown2020language,devlin2018bert},
symbolic music \cite{huang2018music}, video \cite{girdhar2019video,sun2019videobert} and image understanding \cite{dosovitskiy2020image,liu2021swin}. This can be seen
in the light of a broader trend, where a single neural network architecture generalizes across many domains of data and tasks. Attention mechanisms have also been explored for keyword
spotting [12, 13], but only as an extension to other architectures,
such as convolutional or recurrent neural networks.
Inspired by the strength of the simple Vision Transformer
(ViT) model \cite{dosovitskiy2020image} in computer vision and by the techniques that improves its data-efficiency, \cite{berg2021keyword} (As figure \ref{fig:kwt}) proposes an adaptation of this architecture for keyword spotting and find that it matches or outperforms existing models on the Google Speech Commands dataset without additional data. \\

\noindent For user-defined keywords, large datasets are not available since we cannot ask the users to provide many examples. So it can be treated as a few-shot learning problem. Self-supervised representation learning (SSL) methodologies can help by allowing finetuning and pre-learning based on both small and big volumes of labeled and unlabeled data, respectively.
SSL methods promise a single universal model that would benefit a wide variety of tasks and domains. Such
methods have shown success in natural language processing and
computer vision domains, achieving new levels of performance
while reducing the number of labels required for many downstream
scenarios.  Recently, \cite{kao2022efficiency} compares several widely used SSL models to answer which pre-trained model is the best for KWS of few-shot learning. Their result shows that HuBERT \cite{hsu2021hubert} the best result and is robust to the changes of few-shot examples.
\cite{seo2021wav2kws} incorporate Wav2Vec 2.0 \cite{baevski2020wav2vec} , a SSL model, into their KWS models. And \cite{seo2021wav2kws} achieved the state of the art performance that year. \\

\begin{figure}[htp]
  \centering
  \includegraphics[width=0.618\linewidth]{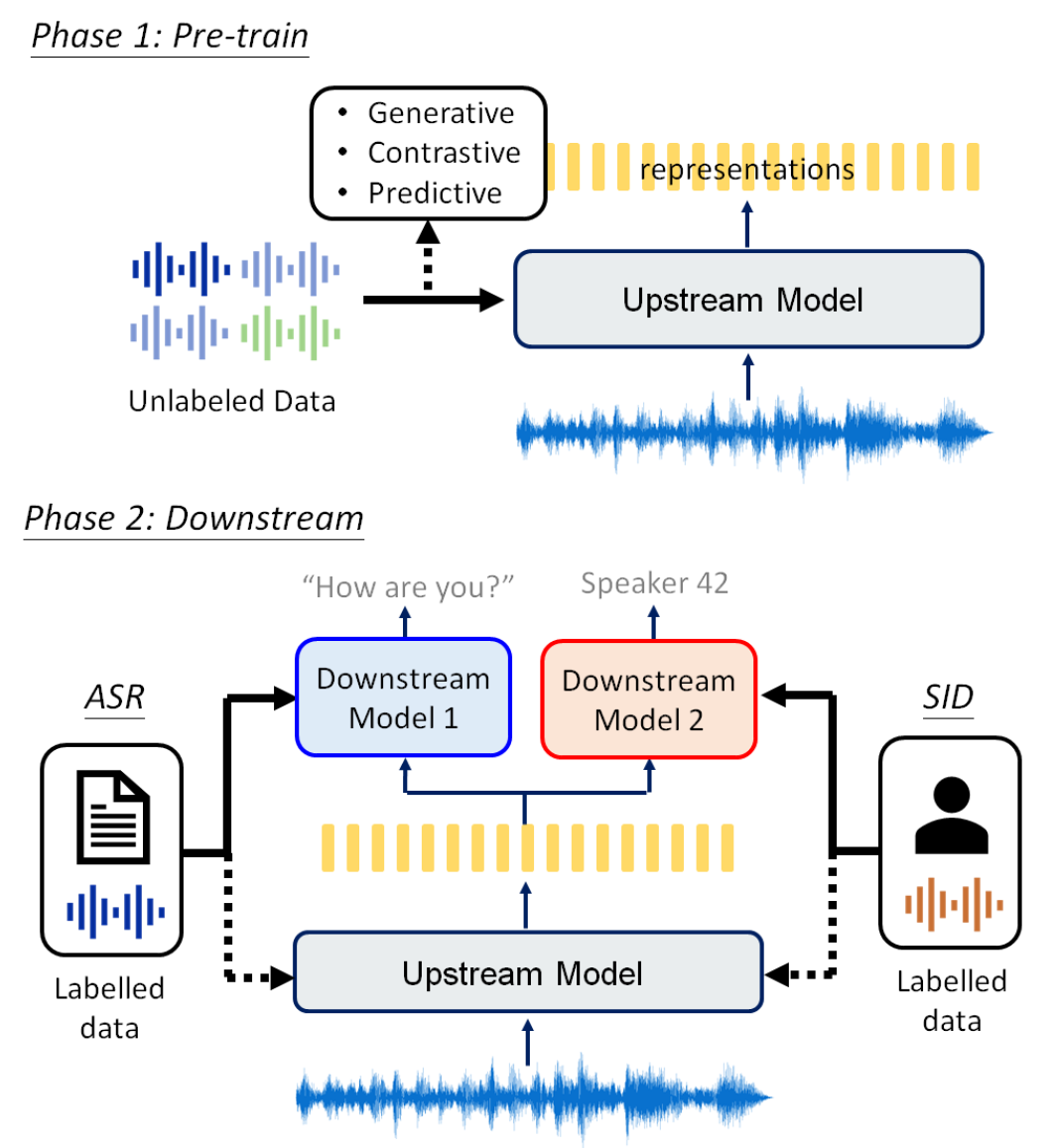}
  \caption{\textit{Framework for using self-supervised representation
learning in downstream applications from \cite{mohamed2022self}.}}
  \label{fig:ssl}
\end{figure}

\noindent We can expect that self-supervised learning of KWS will continue to be a hot topic in the future despite all the progress made. Although there are some research about using self-supervised learning to empower the KWS. We still find out some leaved issues waiting for the solution in next section. How to more efficiently empower the keyword spotting using SSL technologies with unlabelled data is a significant and challenging research task that will be a potential direction for future work. \\

\subsection{Keyword Spotting Dataset}
\begin{figure}[htp]
  \centering
  \includegraphics[width=0.8\linewidth]{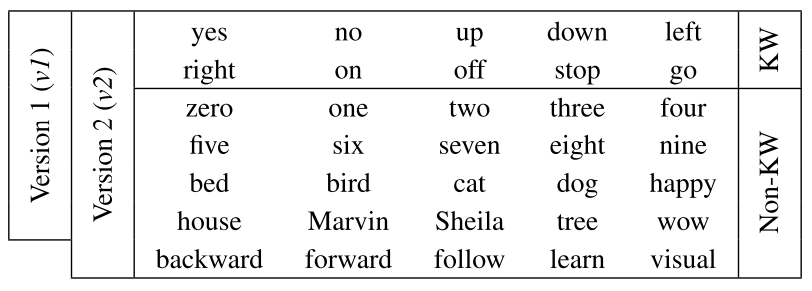}
  \caption{\textit{List of the words from \cite{warden2018speech} included in the Google Speech Commands Dataset v1 (first six rows) and v2 (all the rows). Words are broken down by the standardized 10 keywords (first two rows) and non-keywords (last five rows)}}
  \label{fig:gsc}
\end{figure}

\subsubsection{Google Speech Commands}
\noindent The publicly available Google Speech Commands Dataset \cite{warden2018speech} has become the most famous open benchmark for (deep) KWS development and evaluation. This crowd-sourced database was captured at a sampling rate of \num{16} kHz by means of phone and laptop microphones, being, to some extent, noisy.  Recorded by \num{1881} speakers, this first
version consists of \num{64727} one-second (or less) long speech
segments covering one word each out of \num{30} possible different
words. \\

\noindent The main difference between the first version and
the second version —which was made publicly available in
2018— is that the latter incorporates \num{5} more words (i.e.,
a total of \num{35} words), more speech segments, \num{105829}, and
more speakers, \num{2618}. Figure \ref{fig:gsc} lists the words included in the Google Speech Commands Dataset v1 (first six rows) and v2
(all the rows). To facilitate KWS technology reproducibility and comparison, this benchmark also standardizes the
training, development and test sets, as well as other crucial
aspects of the experimental framework, including a training
data augmentation procedure involving background noises.

\subsubsection{Multi-model Information based Speech Processing (MISP) Challenge}

This dataset \cite{zhou2022audio} considers the following scenario: several people are chatting while watching TV in the living room and they can interact with a smart speaker/TV. In the schematic diagram, six speakers are chatting while multiple devices are used to record the audio and video in parallel. There are some variables that can have an influence on the conversation and/or the collected audio and video that is taking place
in the real living room, for example, the TV can be turned on/off,
the conversation can happen during the day or night, etc. Moreover,
by observing the real conversations taking place in the real living
room, we found that speakers could be divided into several groups
to discuss different topics. This is a common natural conversation
phenomenon. Compared with the situation when all speakers are
discussing the same topic, the grouping results in higher overlap ratios in the audio. We control the above variables to cover as many
real scenes as possible during recording. \\

\begin{figure}[htp]
  \centering
  \includegraphics[width=0.8\linewidth]{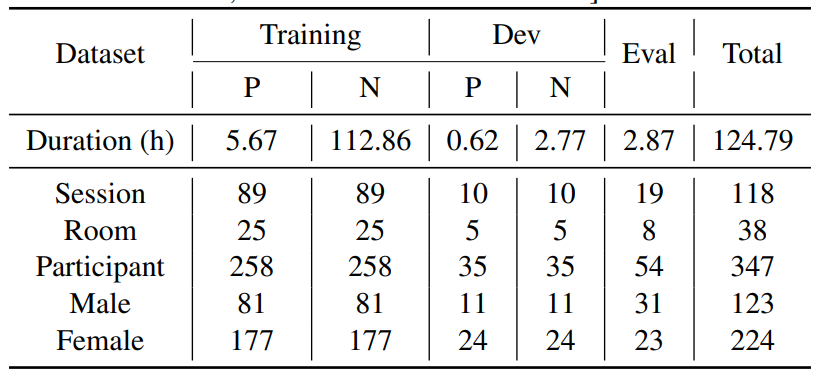}
  \caption{\textit{Overview of the MISP2021-AVWWS corpus. [P: for presence of wake word, N: for absence of wake word]}}
  \label{fig:misp}
\end{figure}

\noindent Three types of recording devices were used. The type of recording
device was dependent on its distance to the speaker. The far recording device is a linear microphone array of 6 sample-synchronised omnidirectional microphones, which is placed 3-5m away from the
speaker. The distance between adjacent microphones is 35 mm. The
far linear microphone array is recorded onto a laptop computer. At
a position 1-1.5 m away from the speaker, we placed a linear microphone array of 2 sample-synchronised omnidirectional microphones.
The distance between adjacent microphones is 92mm. To facilitate
transcription, each speaker wore a high-fidelity microphone, on the
middle of chin. The audio from the middle linear microphone array
and each near high-fidelity microphone was recorded via a sound
board. \\

\noindent The database used for task 1 contains 124.79 hours of audio-visual
data. Figure \ref{fig:misp} shows the division of the audio-visual data into a
training, development, and evaluation set and indicates details regarding the number of sessions, the type of room, and the number
of male/female speakers. The wake word is “Xiao T Xiao T”. The
data set includes 118 sessions. The number of speakers within one
conversation session ranges from 1 to 6. The total number of speakers in the data set is 347. All speakers are native Chinese speaking
Mandarin without strong accents. Various conversation topics were
recommended during recording. Due to the final ranking only lies on
the results of the far recordings, the evaluation set only contains the
recordings from the far devices, but the middle and near recordings
are avail in the training and development sets. Some real noise data
is also provided.

\newpage

\section{Environmental Sound Classification}
\label{sec:esc}
\subsection{Definition and Background}
Environmental sound classification (ESC) aims to categorize audio recordings into pre-defined environmental sound classes \cite{piczak2015environmental}. The set of target sounds for a detection task are specific to each application, but in the case of a general-purpose sound event detection system the target sounds are environmental sounds such as bird singing, car passing by, and footsteps. In the literature, these are sometimes referred to as \textit{nonspeech} and \textit{non-music} sounds \cite{gygi2007environmental}, to differentiate the field of
environmental sound analysis from more established speech or
music analysis tasks. The sound event detection task also has
a different purpose than the typical speech or music analysis
tasks, because perception of speech, music, and environmental
sounds is also different: while musical listening focuses on the
aesthetic qualities of the sound, and speech perception focuses on the linguistic or paralinguistic information, everyday listening is directed towards identification of the sound sources \cite{gaver1993world}. \\

\noindent Recently, on-device environmental sound classification \cite{Singh2022, singh2022passive, choi2022temporal} has attracted increasing research interest, as shown in Task 1 of Detection and Classification of Acoustic Scenes and Events (DCASE) 2022 Challenge: ``Low-Complexity Acoustic Scene Classification" \cite{martin2022low}. Such a sound classification system with low computation-complexity can be deployed on mobile and embedded platform for many real-world audio applications, such as acoustic surveillance \cite{radhakrishnan2005audio}, bio-acoustic monitoring \cite{Liu2022a} and multimedia indexing \cite{kiranyaz2006generic}. \\

\noindent The number of possible sound classes is unlimited,
since any object or being may produce a sound as a naturally occurring event.\\

\subsection{Environmental Sound Classification Approach}

Deep neural networks have brought tremendous improvement in many domains such as image classification and speech
recognition, and are now also the dominant approach in environmental sound analysis and classification, as observed in the
recent years \cite{mesaros2017detection,mesaros2019sound}. Their main drawback is that they require
large amounts of data for training. This need for large datasets
is a problem for sound event detection because the domain
still lacks large datasets of strongly-labeled data. Advanced
training strategies involving weak labels and transfer learning
are providing suitable solutions to cope with shortcomings in
the data, but the general system architectures often do not
change dramatically.\\

\subsubsection{Convolutional Neural Network:}
Traditional CNN architectures use multiple blocks of successive convolution and pooling operations for feature learning and down-sampling along the time and feature dimensions, respectively.
As an alternative, Ren et al. used atrous CNNs, which are based on dilated convolutional kernels \cite{ren2019attention}. Such kernels allow achieving a comparable receptive field size without intermediate pooling operation.
Koutine et al. showed that ESC systems can be improved if the receptive field is regularized by
restricting its size \cite{koutini2019cp}.\\

\noindent In most CNN based architectures, only the activations of the last convolutional layer are connected
to the final classification layers. As an alternative, Yang et al. followed a multi-scale feature approach
and further processed the activations from all intermediate feature maps \cite{yang2018acoustic}. Additionally, the authors
used the Xception network architecture, where the convolution operation is split into a depthwise
(spatial) convolution and a pointwise (channel) convolution to reduce the number of trainable
parameters. A related approach is to factorize two-dimensional convolutions into two one-dimensional
kernels to model the transient and long-term characteristics of sounds separately \cite{cho2019acoustic}. The influence
of different symmetric and asymmetric kernel shapes were systematically evaluated by Wang et al. \cite{wu2017asymmetrie}.\\

\noindent Several extensions to the common CNN architecture were proposed to improve the feature learning.
Basbug and Sert adapted the spatial pyramid pooling strategy from computer vision, where feature maps
are pooled and combined on different spatial resolutions \cite{basbug2019acoustic}. Marchi et al. added
the first and second order time derivative of spectrogram based features as additional input channels in
order to facilitate detecting transient short-term events that have a rapid increase in magnitude \cite{Marchi2016}. \\

\begin{figure}[htp]
  \centering
  \includegraphics[width=0.8\linewidth]{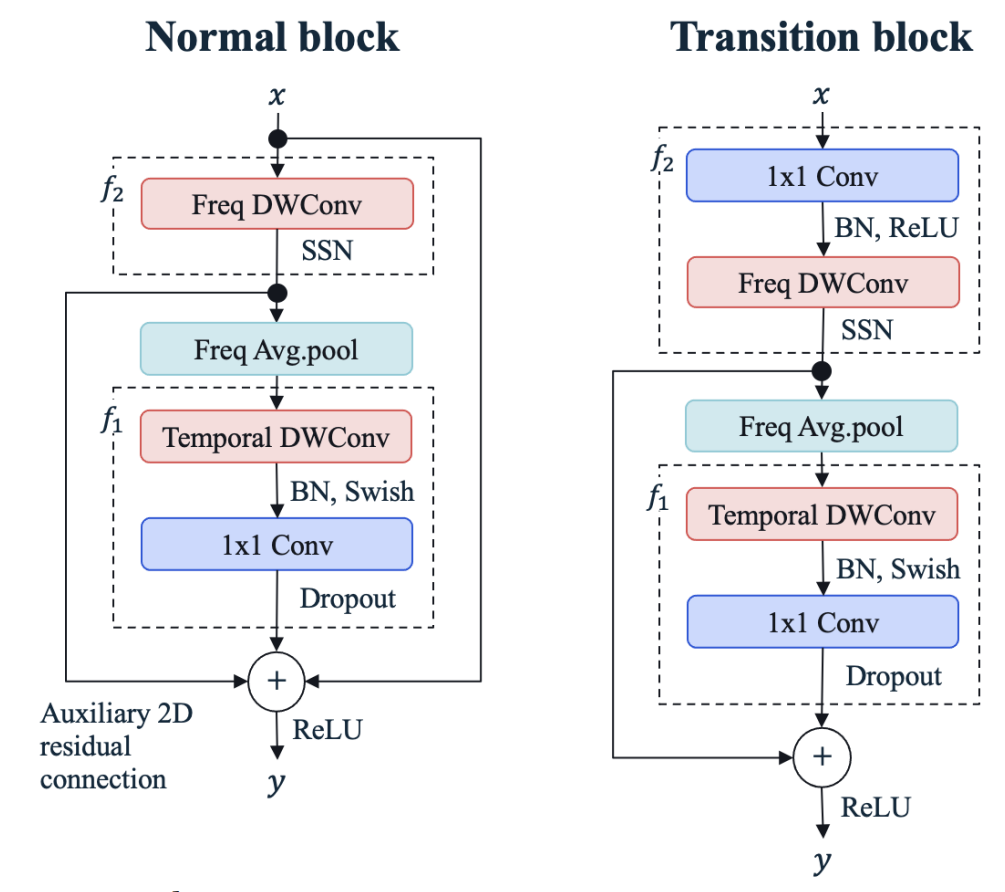}
  \caption{\textit{: BCResBlock from \cite{kim2021broadcasted}. The BC-ResNet block contains a frequency-depthwise convolution with a SubSpectralNorm. Then the feature is averaged by frequency followed by temporal-depthwise separable convolution. Temporal feature is broadcasted to 2D features at residual connection. In a transition block, they have an additional 1x1 convolution on the front to change the number of channel without identity
shortcut.}}
  \label{fig:bcblock}
\end{figure}

\begin{figure}[htp]
  \centering
  \includegraphics[width=0.8\linewidth]{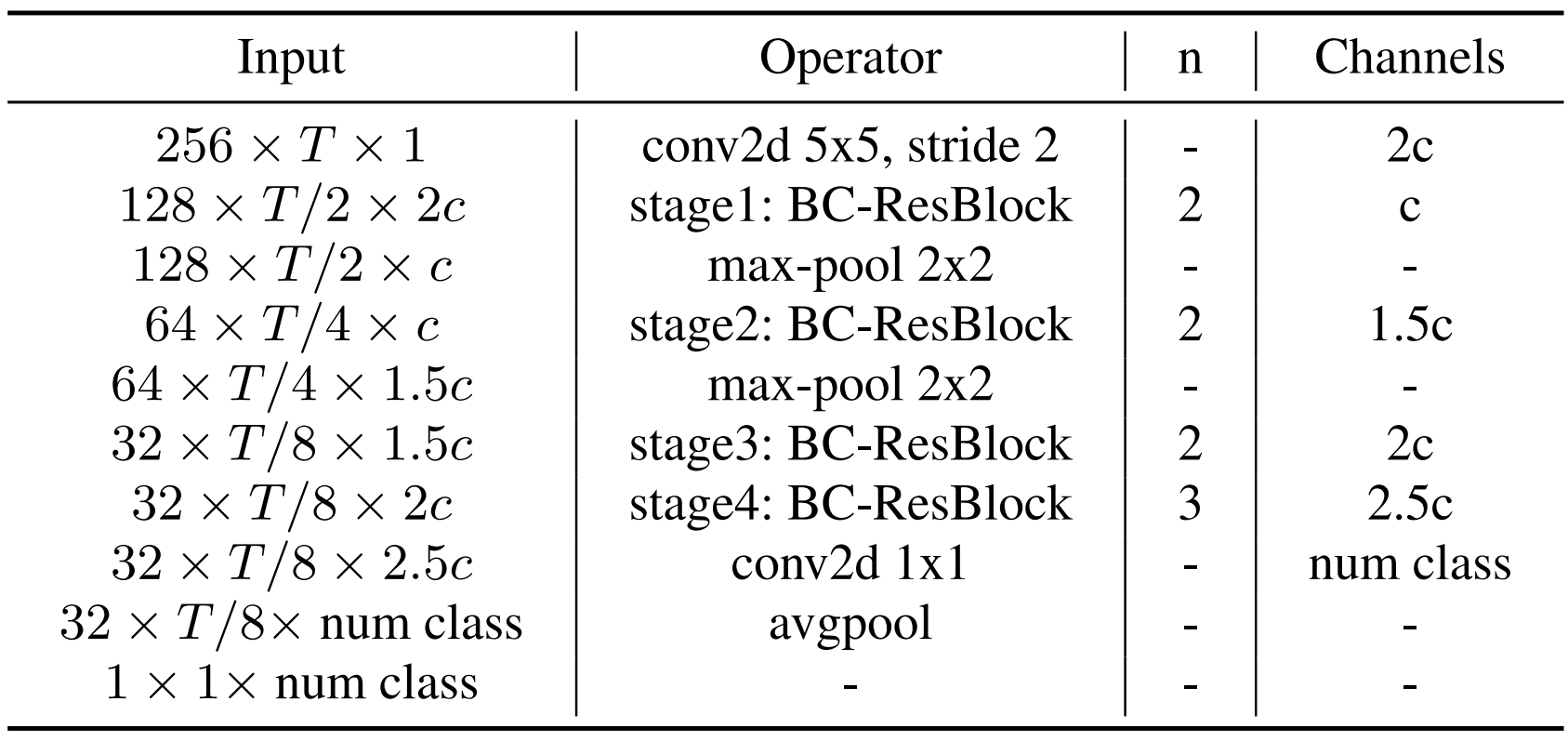}
  \caption{\textit{: BC-ResNet-Mod from \cite{Kim2021b}. Each row is a sequence of one or more identical modules repeated $n$ times with input shape of frequency by time by channel and total time step $T$.}}
  \label{fig:bcmod}
\end{figure}

\noindent In recent years, a number of researches have been proposed for more efficient and high-performance Environmental Sound Classification. Kim et al. proposed methods \cite{Kim2021b} to leverage the generalization capabilities of unseen devices while maintaining the model’s performance in lightweight models. To design a low-complexity network in terms of the number of
parameters, they used a Broadcasting-residual network (BC-ResNet)
\cite{kim2021broadcasted}, a baseline architecture that uses 1D and 2D CNN features together for better efficiency. While the \cite{kim2021broadcasted} targets human voice, we aim to classify the audio scenes. Therefore, we make two modifications to the network, i.e., limit the receptive field and use max-pool instead of dilation, to adapt to the differences in input domains. The proposed architecture is shown in Figure \ref{fig:bcmod}, a fully CNN named modified BC-ResNet (BC-ResNet-Mod). The model has 5x5 convolution on the front with a 2x2 stride for downsampling followed by BC-ResBlocks \cite{kim2021broadcasted}. With a total of 9 BC-ResBlocks as \ref{fig:bcblock} and two maxpool layers, the receptive field size is 109x109. They also do the last 1x1 convolution before global average pooling that the model classifies each receptive field separately and ensembles them by averaging. BC-ResNets use dilation in temporal dimension to obtain a
larger receptive field while maintaining temporal resolution across
the network. And they observe that time resolution does not need to be fully kept in the audio scene domain, and instead of dilation, they insert max-pool layers in the middle of the network.\\

\noindent BC-ResNet-Mod achieve two goals; 1) efficient design in terms of the number of parameters and 2) adapting to device imbalanced dataset. Therefore, for our experiments, we use BC-ResNet-Mod-4, which increases the input channel dimension to 80 before extracting spectral and temporal features.

\subsubsection{Feedforward Neural Network:}
Feedforward neural networks (FNN) are used in several ESC algorithms. Bisot et al. used an FNN architecture to concatenate features from an NMF decomposition and a constant-Q transform
of the audio signal \cite{Bisot2017NonnegativeFL}. Takahashi et al. combined an FNN with multiple Gaussian mixture model (GMM) classifiers to model the individual acoustic scenes \cite{8282314}.

\subsubsection{Convolutional Recurrent Neural Network:}
A general-purpose network architecture for sound event detection is the convolutional recurrent neural network (CRNN), containing convolutional and recurrent layers that have specific roles \cite{cakir2017convolutional}. The convolutional layers act as feature extractors, aiming to learn discriminative features through the consecutive convolutions and non-linear transformations applied to the time-frequency representation presented at the input of the
network. The recurrent layers have the role of learning the
temporal dependencies in the sequence of features presented at
their input. 

\subsubsection{Transformer:}
In order to better capture long-range
global context, a recent trend is to add a self-attention mechanism on top of the CNN. Such CNN-attention hybrid models have achieved state-of-the-art (SOTA) results for many audio classification tasks such as audio event classification \cite{kong2020panns,gong2021psla},
spoken keyword spotting \cite{rybakov2020streaming}, and emotion recognition \cite{li2018attention}.
However, motivated by the success of purely attention-based
models in the vision domain \cite{dosovitskiy2020image,liu2021swin}, it is reasonable to ask
whether a CNN is still essential for audio classification. \\

\begin{figure}[htp]
  \centering
  \includegraphics[width=0.618\linewidth]{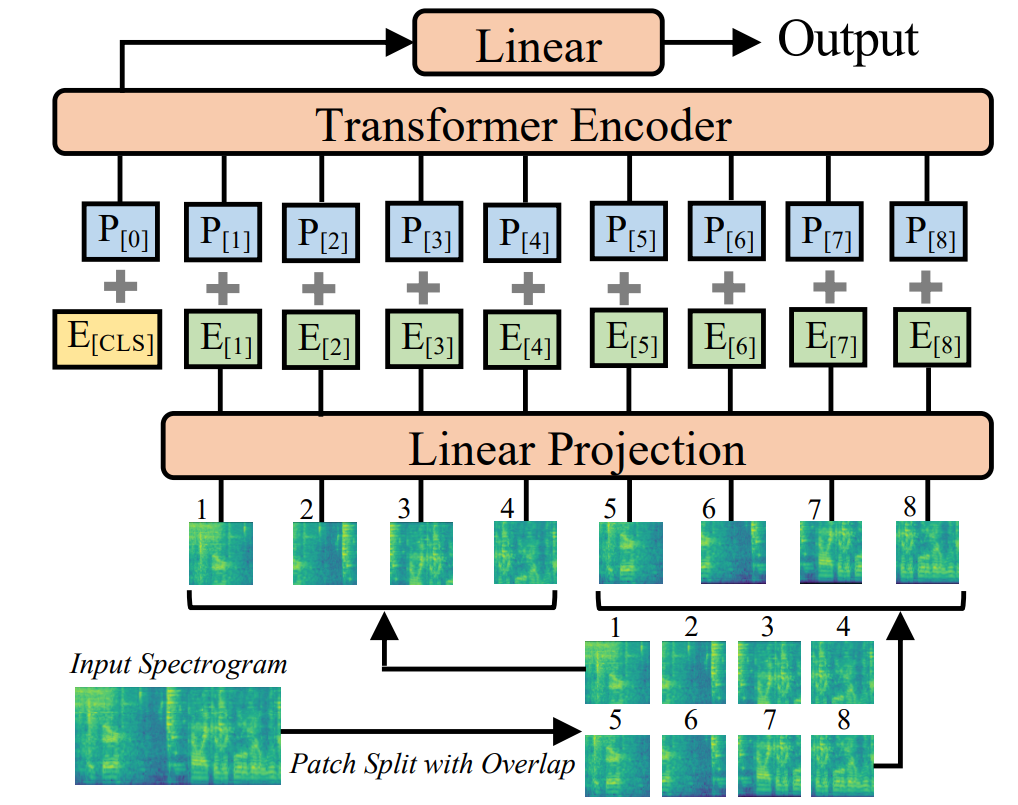}
  \caption{\textit{Audio Spectrogram Transformer (AST)
architecture from \cite{gong2021ast}. The 2D audio spectrogram is split into a sequence of 16×16 patches with overlap, and then linearly projected to a sequence of 1-D patch embeddings. Each patch embedding is added with a learnable positional embedding. An additional
classification token is prepended to the sequence. The output
embedding is input to a Transformer, and the output of the classification token is used for classification with a linear layer}}
  \label{fig:ast}
\end{figure}

\begin{figure}[htp]
  \centering
  \includegraphics[width=0.618\linewidth]{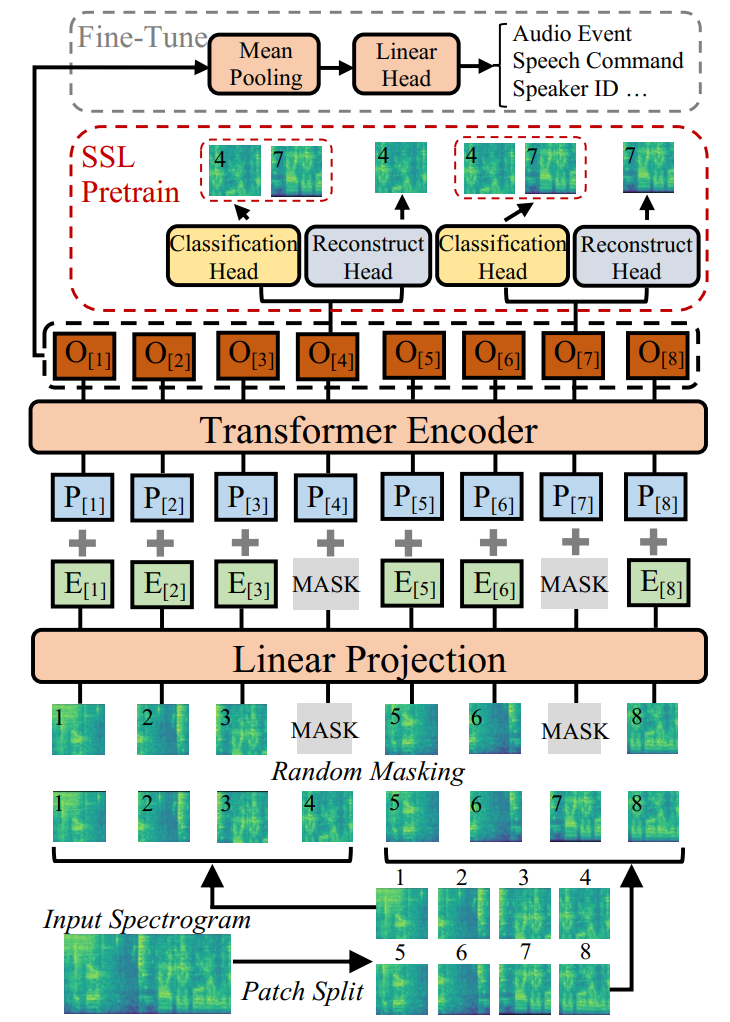}
  \caption{\textit{The self-supervised AST from \cite{gong2022ssast}. The 2D audio spectrogram is split into a sequence of 16×16 patches without overlap, and then linearly projected to a sequence of 1-D patch embeddings $E$. Each patch embedding is added with a learnable positional embedding P and then input to the Transformer encoder. The output of the Transformer $O$ is used as the spectrogram patch representation. During self-supervised pretraining, they randomly mask a portion of spectrogram patches and ask the model to 1) find the correct patch at each masked position from all masked patches; and 2) reconstruct the masked patch. The two pretext tasks aim to force the AST model to learn both the temporal and frequency structure of the audio data. During fine-tuning, they apply a mean pooling over all patch representation $\{O\}$ and use a linear head for classification.}}
  \label{fig:ssast}
\end{figure}

\noindent Gong et al. introduced the Audio Spectrogram Transformer (AST) \cite{gong2021ast}, a convolution-free, purely attention-based model that is directly applied to an audio spectrogram and can capture long-range global context even in the lowest layers. AST (Figure \ref{fig:ast}) outperforms state-of-the-art systems on a variety of audio classification tasks and datasets. \\

\noindent While annotating audio and speech data is expensive, they explores Self-Supervised AST (SSAST) that leverages unlabeled data to alleviate the data requirement problem. In \cite{gong2022ssast},  Gong et al. present a novel joint discriminative and generative Masked Spectrogram Patch Modeling (MSPM) based self-supervised learning (SSL) framework that can significantly improve AST performance with limited labeled data. To the best of our knowledge, MSPM is the first patch-based self-supervised learning framework in the audio and speech domains, and SSAST is the first self-supervised pure self-attention based audio classification model. They also show that pretraining with both speech and audio
datasets noticeably improves the models’ generalization
ability, and leads to better performance than pretraining
with dataset from a single domain. As a consequence,
the SSAST model performs well on both speech and audio downstream tasks.

\newpage

\subsection{Sound Classification Dataset}
\subsubsection{Audio set}
Audio set \cite{gemmeke2017audio} is a large scale dataset of manually-annotated audio events that endeavors to
bridge the gap in data availability between image and audio research. Using a carefully structured hierarchical ontology of 632
audio classes guided by the literature and manual curation, we collect data from human labelers to probe the presence of specific audio classes in 10 second segments of YouTube videos. Segments are proposed for labeling using searches based on metadata, context (e.g.,links), and content analysis. The resulting dataset includes 1,789,621 segments (4,971 hours), comprising at least 100 instances for 485 audio event categories.
\begin{figure}[htp]
  \centering
  \includegraphics[width=0.5\linewidth]{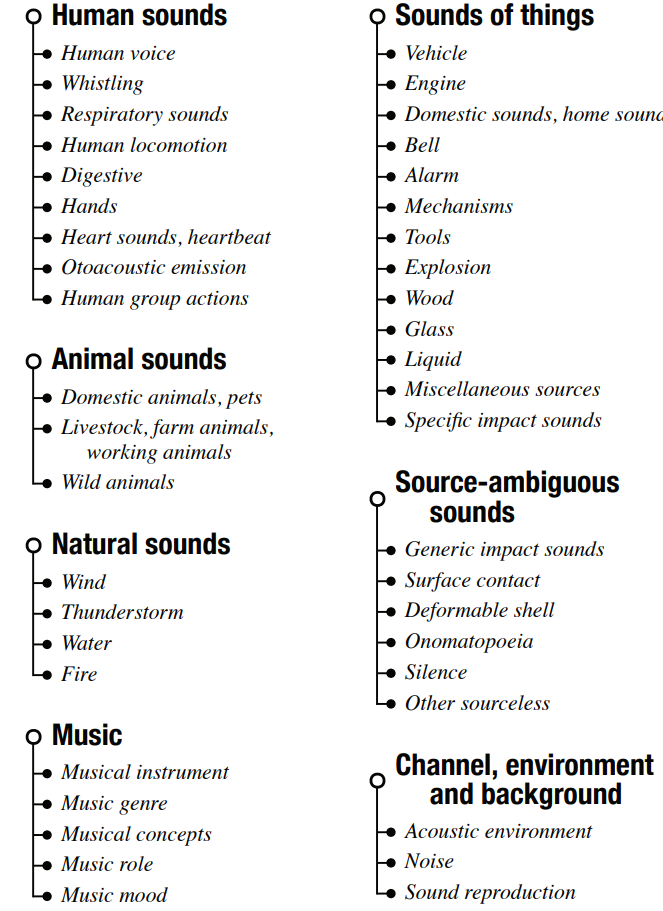}
  \caption{\textit{The categories from \cite{gemmeke2017audio}}}
  \label{fig:audioset}
\end{figure}
\subsubsection{TAU Urban Acoustic Scenes 2019}
The TAU Urban Acoustic Scenes 2019 dataset \cite{tau2019}, consisting of recordings from the following acoustic scenes: airport, indoor shopping mall, metro station, pedestrian street, public square, street with medium level of traffic, travelling by tram, travelling by bus, travelling by underground metro, and urban park.\\

\noindent The dataset used for the task is an extension of the TUT 2018
Urban Acoustic Scenes dataset, recorded in multiple cities in Europe. TUT 2018 Urban Acoustic Scenes dataset contains recordings
from Barcelona, Helsinki, London, Paris, Stockholm and Vienna, to
which TAU 2019 Urban Acoustic Scenes dataset adds Lisbon, Amsterdam, Lyon, Madrid, Milan, and Prague. The recordings were
done with four devices simultaneously.\\
\subsubsection{ESC-50}
The ESC-50 \cite{piczak2015esc} dataset consists of 2,000 labeled environmental recordings equally balanced between 50 classes (40 clips per class). For convenience, they are grouped in 5 loosely
defined major categories (10 classes per category):\\

\begin{itemize}
    \item  animal sounds,
\item natural soundscapes and water sounds,
\item human (non-speech) sounds,
\item  interior/domestic sounds,
\item exterior/urban noises.
\end{itemize}
\begin{figure}[htp]
  \centering
  \includegraphics[width=\linewidth]{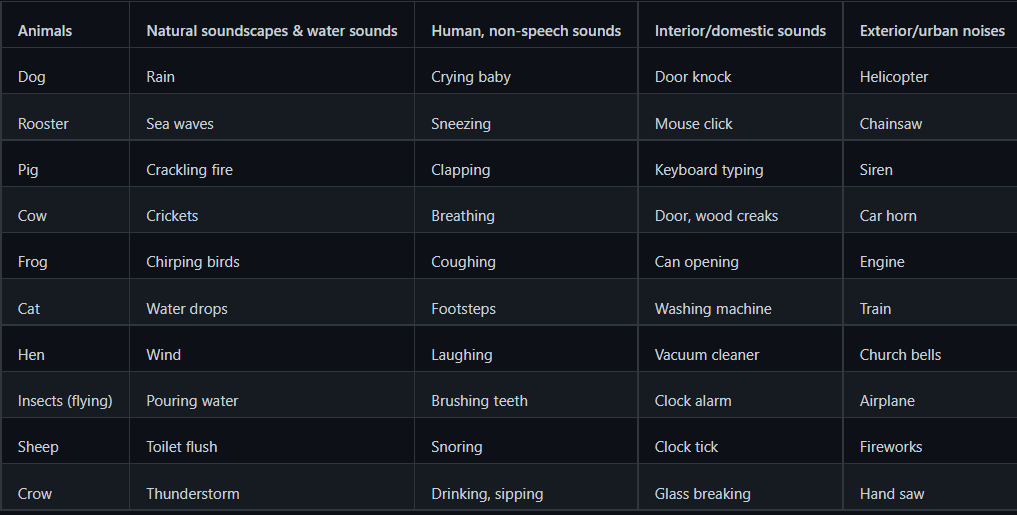}
  \caption{\textit{The categories from \cite{piczak2015esc}}}
  \label{fig:esc50}
\end{figure}

\noindent The dataset provides an exposure to a variety of sound
sources - some very common (laughter, cat meowing, dog barking), some quite distinct (glass breaking, brushing teeth) and
then some where the differences are more nuanced (helicopter
and airplane noise).

\newpage

\section{Continual Learning}
\label{sec:cl}
\subsection{Definition and Background}

Continual learning aims to develop artificial intelligence systems that can continuously learn to solve new tasks from new data while retaining knowledge learned from previously learned tasks \cite{masana2020class,rebuffi2017icarl}. In most continual learning (CL) scenarios, learners are presented with tasks in a series of illustrated training sessions. During that time, only data from a single task are used for training. After each training, the learner should be able to perform all previously seen tasks on unseen data. The biological inspiration for this learning model is evident as it reflects how humans acquire and integrate new knowledge. When presented with a new learning task, it leverages previous knowledge and integrates newly learned knowledge into previous tasks. \\

\noindent This is in stark contrast to common supervised learning paradigms, where the labeled data for all tasks are jointly available during a single deep network training session. A continuing learner can only access data for one of her tasks at a time while assessing all previous learning tasks. The main challenge of continual learning is learning data from the current task so as not to forget previously learned tasks. The naive fine-tuning method applies very effectively to the domain transfer problem, but the data from the previous task is missing and the resulting classifier cannot classify the data from it. A dramatic drop in performance on a previously learned task is a phenomenon known as catastrophic forgetting \cite{mccloskey1989catastrophic}. Continual learning aims to prevent \textit{catastrophic forgetting}, while at the same time avoiding the problem of \textit{intransigence} which inhibits adaptation to new tasks. \\

\begin{figure*}[htbp]
  \centering
  \includegraphics[width=0.618\linewidth]{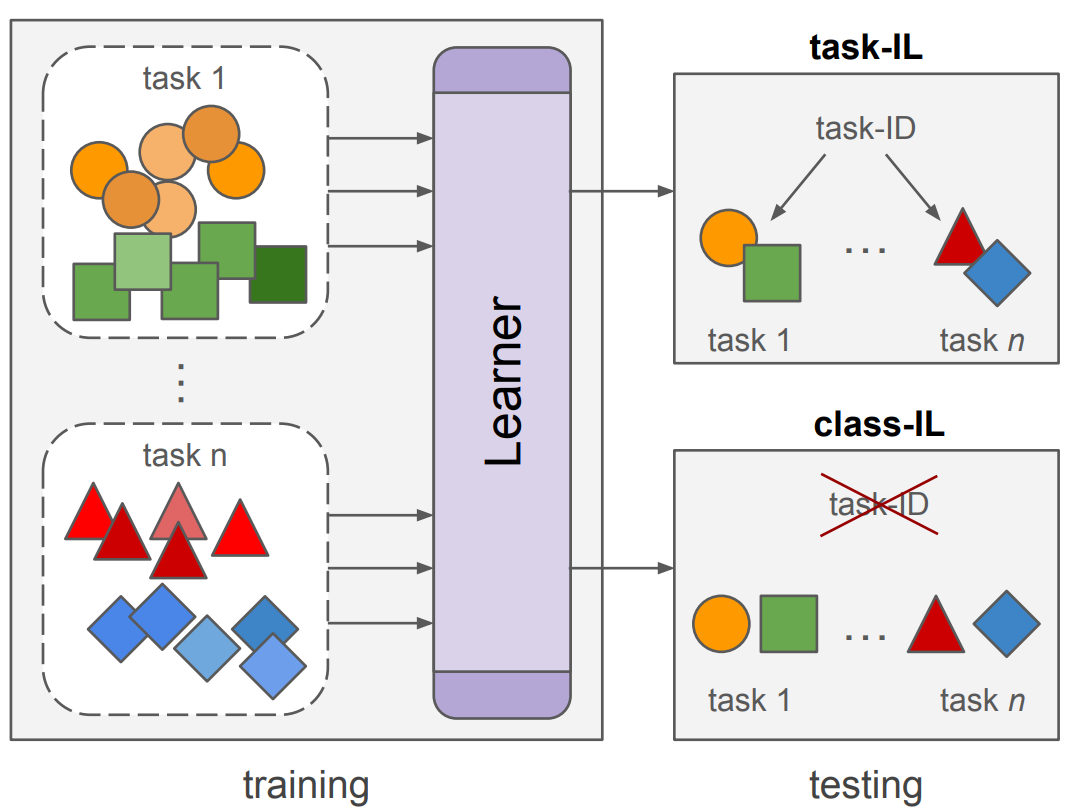}
  \caption{\textit{In continual learning, disjoint tasks are learned sequentially. Task-IL has access to the task-ID during evaluation, while the more challenging setting of class-IL does
not. Class-IL is the subject of this thesis.}}
  \label{fig:cl}
\end{figure*}

\noindent Continual learning divides training into a series of tasks. In any training session, learners can only access data for their current task. Optionally, some methods can take into account small amounts of data saved from previous tasks. Most early approaches considered this situation, called task incremental learning (task IL), where the algorithm had access to the task ID at inference time. This has the obvious advantage that the method does not need to distinguish the class from different tasks. Recently, approaches have been initiated to address the more difficult class-incremental learning (class-IL) scenarios, where learners do not have access to task IDs at inference time and must be able to distinguish between all classes and all classes (see figure \ref{fig:cl}). Scenarios where task IDs are not present at inference time typically include scenarios that gradually increase capacity granularity (for example, detecting more and more object classes in images). Various class-IL techniques have been proposed in the last few years, and this paper also focuses on class-IL techniques in continual learning.\\

\noindent A significant increase in the popularity of CL over the past few years has been driven by the demand for industrial and social applications. Gradual assimilation of knowledge can solve some problems: \\

\begin{itemize}
    \item \textbf{Memory restrictions:} A system that has physical constraints on what data it can store cannot store all the data it can display, so it cannot rely on joint training strategies. Such systems can only store a limited set of examples of tasks to perform and must be learned incrementally. 
    \item \textbf{Data security/privacy restrictions:} Continual learning can provide a solution for systems learning from data that should not be stored permanently. Government laws may restrict customers from storing data in central locations (e.g. for applications on their mobile devices).
   \item \textbf{Sustainable artificial intelligence:} Training deep learning algorithms can be expensive. The carbon footprint of retraining such systems with every new data update is substantial and likely to grow in the coming years. Continual learning provides algorithms that are computationally efficient and only need to process new data when the system is updated.

\end{itemize}

\subsection{Continual Learning Approach}
In this section, we describe several approaches to address
the above mentioned challenges of CL. We divide them
into three main categories: regularization-based methods,
rehearsal-based methods, and bias-correction methods.

\subsubsection{Regularization approaches}
Several approaches use regularization terms together with
the classification loss in order to mitigate catastrophic forgetting. Some regularize on the weights and estimate an
importance metric for each parameter in the network \cite{ewc}, while others focus on the importance of remembering feature representations \cite{li2017learning}.

\subsubsection{Rehearsal approaches}
Rehearsal methods keep a small number of exemplars (exemplar rehearsal), or generate synthetic samples. By
replaying the stored or generated data from previous tasks
rehearsal methods aim to prevent the forgetting of previous
tasks. Most rehearsal methods combine the usage of exemplars to tackle the inter-task confusion with approaches that
deal with other causes of catastrophic forgetting. The usage
of exemplar rehearsal for CL was first proposed in Incremental Classifier and Representation Learning (iCaRL) \cite{rebuffi2017icarl}. This technique has since been applied in the majority of CL methods. In next two sections, we focus on the choices which need to be taken when applying exemplars.

\subsubsection{Bias-correction approaches}

Bias-correction methods aim to address the problem of task-recency bias, which refers to the tendency of incrementally
learned network to be biased towards classes in the most
recently learned task. This is mainly caused by the fact that,
at the end of training, the network has seen many examples
of the classes in the last task but none (or very few in case of
rehearsal) from earlier tasks.  One simple and effective approach to preventing task-recency bias was proposed by Wu et al \cite{wu2019bic}, who call their method \textit{Bias Correction} (BiC). They add an additional layer dedicated to correcting task bias to the network. A training session is divided into two stages. During the first stage they train the new task with the cross-entropy loss. Then they use a split of a very small part of the training data to serve as a validation set during the second phase.
\newpage

%% file: Chapter3/Chapter3.tex

\chapter{Continual Learning for Spoken Keyword Spotting}
In this chapter, we proposes a novel diversity-aware continual learning approach named Rainbow Keywords (RK) to address the issues mentioned above, requiring no task-ID information with fewer parameters. Specifically, the proposed RK approach introduces a diversity-aware sampler to select few but diverse examples from historical and incoming keywords by calculating classification uncertainty. As a result, the model will not forget the prior knowledge when learning new keywords even utilizing limited historical examples. Furthermore, we utilize a mixed-labeled data augmentation to additionally improve the diversity of selected examples for higher performances. Besides, we propose a knowledge distillation loss function to guarantee that the prior knowledge could remain from the limited selected examples. We conduct our experiments on \textit{Google Speech Command} dataset following the setup of prior work \cite{mai2022online,prabhu2020gdumb}. Experimental results show that the proposed RK approach achieves 4.2\% absolute improvement in terms of Average Accuracy over the best baseline with less required memory. The scripts are available on GitHub \footnote{https://github.com/swagshaw/Rainbow-Keywords}.
\section{Related Works}
Prior work \cite{awasthi2021teaching,2021Few} utilize few-shot fine-tuning to adapt KWS models with training data from the target-domain for new scenarios. However, performances on data from the source domain after adaptation could be poor, which is also known as the  \textit{catastrophic forgetting} problem \cite{mccloskey1989catastrophic}. Recent work \cite{huang2022progressive} proposes a progressive continual learning \cite{delange2021continual} strategy for small-footprint keyword spotting to alleviate the catastrophic forgetting problem when adapting the model trained by source-domain data with the target-domain data. The limitations of such an approach are two-fold. First, the approach requires the task-ID as auxiliary information to learn the knowledge of different tasks, which is not always available in practice. Second, the storage volume occupied by the model will increase with the higher task numbers. The storage volume will be unaffordable for light edge devices. 

\section{Method}

With online keyword spotting systems, we assume that the model should identify all keywords in a series of tasks without catastrophic forgetting. For each task $\tau_t$, we have input pairs $(x_t,y_t)$, where $x_t$ denote audio utterances and $y_t$ are keyword labels. We aim to minimize a cross-entropy loss \cite{masana2020class} of all keywords $N^t$ up to the current task $\tau_t$ formulated as :
\begin{equation}
L_{CE}(x,y) = \sum\limits^{N^t}_{i=1} y_i log \frac{exp(o_i)}{\sum^{N^t}_{j=1} exp(o_j)}
\label{eq1}
\end{equation}
where $o$ denotes the output logits of the model in the task $\tau_t$.

\begin{figure*}[t]
  \centering
  \includegraphics[width=\linewidth, height=0.618\linewidth]{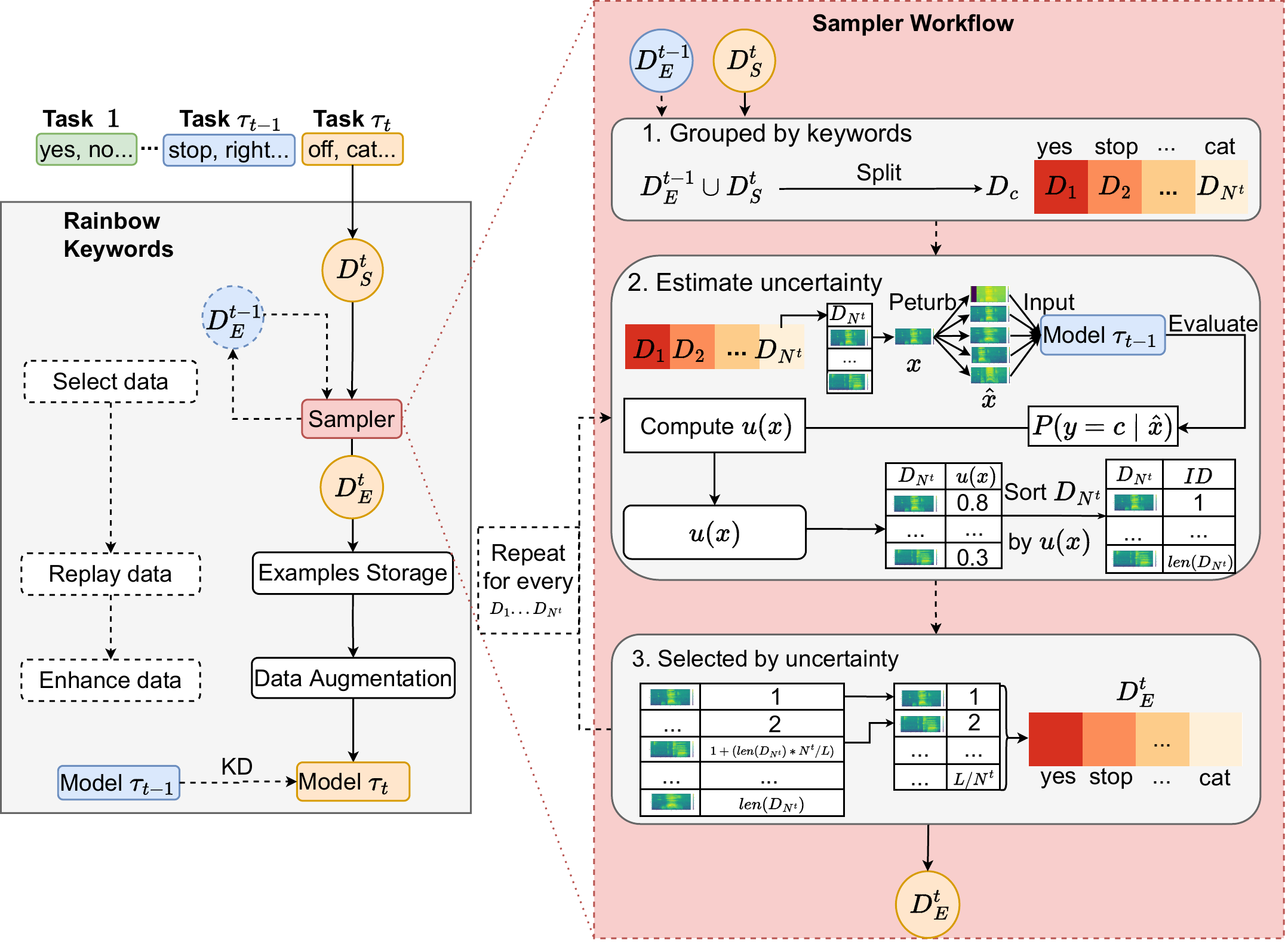}
  \caption{\textit{Block diagram of the proposed Rainbow Keywords approach. Specifically, $D^t_S$ denotes incoming audio stream data of the task $\tau_t$. $D^t_E$ and $D^{t-1}_E$ denote the examples of the task $\tau_t$ and $\tau_{t-1}$, respectively. We group $D^{t-1}_E \cup D^t_S$ into subsets as $D_c, c=1...N^t$ by unique keywords, where $N^t$ denotes the total numbers of unique keywords in $D^{t-1}_E \cup D^t_S$ set. $x$, $\hat{x}$ and $K$ present each sample in $D_c$, the five perturbations of $x$ and the five perturbation strategies. ``Compute $u(x)$" is to compute $u(x)$ by Eq.\ref{ux}.}}
  \label{fig:workflow}
\end{figure*}

\subsection{Diversity-Aware Sampler}
The diversity-aware sampler aims to select \textit{diverse examples} to manage memory efficiently. Such diverse examples are defined by the relative location of each example in feature space, which are estimated by the uncertainty of the sample through the inference by the classification model \cite{bang2021rainbow}. The required three steps are shown in the right red box of Figure \ref{fig:workflow}: \\


\noindent \textbf{Split by keywords:} The first step gathers the historical examples $D^{t-1}_E$ and incoming data $D^t_S$, and groups them into subsets as $D_c, c=1...N^t$ by unique keywords, where $N^t$ denotes the total numbers of unique keywords in $D^{t-1}_E \cup D^t_S$ set. \\


\noindent \textbf{Estimate uncertainty:} The second step estimates the uncertainty of each sample $x$ in $D_c$ by Monte-Carlo (MC) method \cite{gal2016dropout}, which is defined in Equation \ref{eq3}. \\
\begin{equation}
\label{eq3}
    P(y = c \mid x) = \int p(y=c \mid \hat{x})p(\hat{x}\mid x) d\hat{x}
\end{equation}
where $x$, $\hat{x}$, $y$ denote each audio utterance of keyword $D_{N^t}$, the five perturbations of $x$, and the label of $x$. Therefore, the uncertainty of the audio utterance $x$ is formulated as $u(x)$: \\
\begin{equation}
\label{ux}
    u(x) \approx 1- \frac{1}{K} \sum\limits^K_{k=1} P(y = c\mid \hat{x}_k)
\end{equation}

\noindent where $K$ presents five perturbation strategies, including Clipping Distortion \cite{park2019specaugment}, TimeMask \cite{park2019specaugment}, Shift \cite{ko2015audio}, PitchShift \cite{ko2015audio} and FrequencyMask \cite{park2019specaugment}. The larger $u(x)$ indicates that the less confidence of model to predict the perturbations.\\

\noindent \textbf{Select by uncertainty:} The third step selects $L$ examples from $D_c$ descending by uncertainty $u(x)$ with the step size of $len(D_c)*N^t/L$. As a result, the most diversity examples are included in $D_{E}^{t}$. Only these examples are available for training. \\

\noindent Algorithm \ref{alg:algorithm1} summarizes our proposed diversity aware memory update algorithm. We position the memory size $L$ to determine the max number of samples can be reserved in the device. We assign the same amount of memory slots ($k_c$) over the 'seen' classes ($N^t$). After assigning the exemplars to the memory slots, we compute the uncertainties for both streamed samples ($D^t_S$) and historical memory samples ($D^{t-1}_E$) at task $\tau_t$, then sort all these samples ($D_c$) by their uncertainties. From the sorted list, we select samples with an interval $\mid D_c\mid/k_c$ to secure the diversity. 

\begin{algorithm}[t]
\caption{Diversity-Aware Memory Update.}
\label{alg:algorithm1}
\KwIn{$L$ denotes memory size, $N^t$ denotes the number of seen classes until task $\tau_t$, $D^t_S$ denotes incoming data
at task $\tau_t$, $D^{t-1}_E$ denotes historical examples after task $\tau_{t-1}$}
\KwOut{$D^t_E$ examples after learning task $\tau_t$.}
\BlankLine
Initialize $D^t_E = \{\}$;\\
$k_c = floor(L/N^t)$;\\
\For{class $c = 1,2,...,N^t$}{
$D_c = \{(x,y)\mid y=c,(x,y)\in D^t_S \cup D^{t-1}_E \}$\\
Sort $D_c$ by $u(x)$ computed by Eq.\ref{ux}\\
\For{$j = 1,2,...,k_c$}{$i = j*\mid D_c\mid/k_c$\\
$D^t_E += D_c[i]$\\
}
}
\end{algorithm}

\subsection{Data Augmentation}
As the examples in $D_{E}^{t}$ are few due to memory limitation, we apply the data augmentation to further increase the diversity of $D_{E}^{t}$. Specifically, we randomly mix two audio utterances to increase the amounts of training data without extra storage.

\subsection{Knowledge Distillation Loss}
Recent studies \cite{hinton2015distilling, wu2019bic,rebuffi2017icarl} show that Knowledge Distillation (KD) is effective for transferring knowledge between teacher-student models. Inspired by such theory, we consider the model of task $\tau_{t-1}$ as the teacher model and the model of task $\tau_{t}$ as the student model. We propose a knowledge distillation loss to preserve the prior knowledge from the teacher for the student model to avoid catastrophic forgetting, which is formulated as:\\

\begin{equation}
\begin{aligned}
\sigma(o_i^t(x);N^{t-1}) & = \frac{exp(o_i^t/T)}{\sum^{N^{t-1}}_{j=1} exp(o_j^t/T)} \\
L_{KD}(o^t(x),o^{t-1}(x)) & = \sum\limits^{N^{t-1}}_{i=1} \sigma(o^{t}_i(x)) log \sigma(o^{t-1}_i(x))
\label{eq7}
\end{aligned}
\end{equation}
where $o^{t-1}(x)$ and $o^t(x)$ denote the output logits of the teacher model and student model, respectively. $N^t$ is all keywords up to the task $\tau_t$. $\sigma$ is the knowledge distillation softmax function parameterized by the temperature $T$. The temperature $T$ is the experiential hyper-parameters of knowledge distillation set as 2.0. As a result, we aim to minimize the total loss of all keywords $N^t$ up to the current task $\tau_{t}$ formulated as:\\
\begin{equation}
\begin{aligned}
L_{total}(x,y) & = \lambda L_{CE}(x,y) + & (1-\lambda) L_{KD}(o^t(x),o^{t-1}(x))
\end{aligned}
\end{equation}
Where $L_{CE}$ is the cross-entropy loss defined in Eq.\ref{eq1}, and $L_{KD}$ is the knowledge distillation loss defined above. $\lambda$ is the experiential hyper-parameters defined as $\sqrt{1-\frac{N^{t-1}}{N^t}}$.

\section{Experiments and Results}
\subsection{Dataset}
We conduct experiments on the \textit{Google Speech Command}
dataset v1 (GSC) \cite{warden2018speech}, which includes 64,727 one-second audio clips with 30 English keywords categories. We utilize 80\% of data for training and 20\% of data for testing. All of the audio clips in GSC are sampled at 16kHz in our experiment. 

\subsection{Experimental Setup}
\subsubsection{Network configuration} 
\begin{figure*}[htbp]
  \centering
  \includegraphics[width=0.20\linewidth]{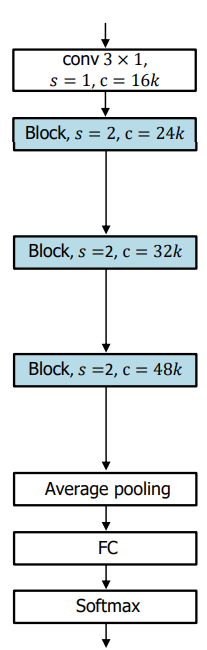}
  \caption{\textit{Architecture for TC-ResNet-8. It utilizes ResNet-8 as the backbone-CNN, respectively. FC denotes fully connected layer. Note that ‘s’, ‘c’, and ‘k’ indicates stride, channel size, and width multiplier, respectively.}}
  \label{fig:tc8}
\end{figure*}

We employ the TC-ResNet-8 \cite{choi2019temporal} as our testbed to evaluate the proposed rainbow keywords approach. It includes a 1-D convolution layer followed by three residual blocks, which consist of 1-D convolution, batch normalization and ReLU active function. Each layer has \{16,24,32,48\} channels (as Figure \ref{fig:tc8}).\\

\noindent We first pre-train the TC-ResNet-8 model on the GSC dataset with 32,000 audio clips, including 15 unique keywords. To evaluate the learning ability of the proposed RK approach, we split the rest data as 5 tasks. Each task includes 3 new unique keywords, which is unseen in previous tasks. To simulate the condition of edge devices, we set the max amount of examples $L$ due to the limited memory in edge devices \cite{mai2022online,prabhu2020gdumb}. \\


\noindent During the training stage, we utilize the Mel-frequency cepstrum coefficients (MFCC = 40) as inputs.

\subsubsection{Reference baselines}
We built eight baselines for comparisons. 
\begin{itemize}
    \item \textbf{Fine-tune training}: adapts the TC-ResNet-8 model for each new task without class incremental learning (CIL) strategies. We consider it as the lower-bound baseline.
    \item \textbf{NR \cite{hsu2018nr}:} is a CIL approach which randomly selects training samples from previous tasks for future training.
    \item \textbf{iCaRL \cite{rebuffi2017icarl}:} is a CIL approach which selects the samples close to the mean of its own class. Then iCaRL utilizes examples for future training.
    \item \textbf{EWC \cite{ewc}:} is a CIL approach which incorporates a quadratic penalty to regularize parameters of model that were important to past tasks. The importance of parameters is approximated by the Fisher Information Matrix.
    \item \textbf{RWalk \cite{rwalk}:} is a CIL approach which improves the EWC. Both Fisher Information Matrix approximation \cite{ewc} and online path integral \cite{zenke2017pi} are fused to calculate the importance for each parameter. RWalk also selects historical examples and utilizes them for future training.
    \item \textbf{BiC \cite{wu2019bic}:} is a recent CIL approach with more attentions. BiC introduces an additional layer to correct task bias of the network. BiC also uses the same sampling method as iCaRL to select historical examples.
    \item \textbf{PCL-KWS \cite{huang2022progressive}:} is a CL approach for Spoken Keyword Spotting. Specifically, the PCL-KWS includes several task-specific sub-networks to memorize the knowledge of the previous keywords. Then, a keyword-aware network scaling mechanism is introduced to reduce the network parameters. The PCL-KWS requires the task-ID to select conspronding sub-networks.
    \item \textbf{Joint training:} trains the TC-ResNet-8 model with the whole dataset, regardless of any constrains. We consider it as the upper-bound baseline.
\end{itemize}

\subsubsection{Metrics}
We report performances in terms of the accuracy and efficiency metrics. The accuracy metrics include \textit{Average Accuracy} (ACC), and \textit{Backward Transfer} (BWT) \cite{lopez2017gradient}. Specifically, the `Average Accuracy' reports an accuracy averaged on all learned tasks after the entire training ends. The “BWT” evaluates accuracy changes on all previous tasks after learning a new task, indicating the forgetting degree. The efficiency metrics include Parameters and Memory. The `Parameter' measures the total parameters of the model in the strategy. The `Memory' indicates the memory requirement of total training data in each task.
\begin{figure*}[!t]
  \centering
  \begin{minipage}[t]{0.49\linewidth}
  \centering
  \includegraphics[width=\linewidth]{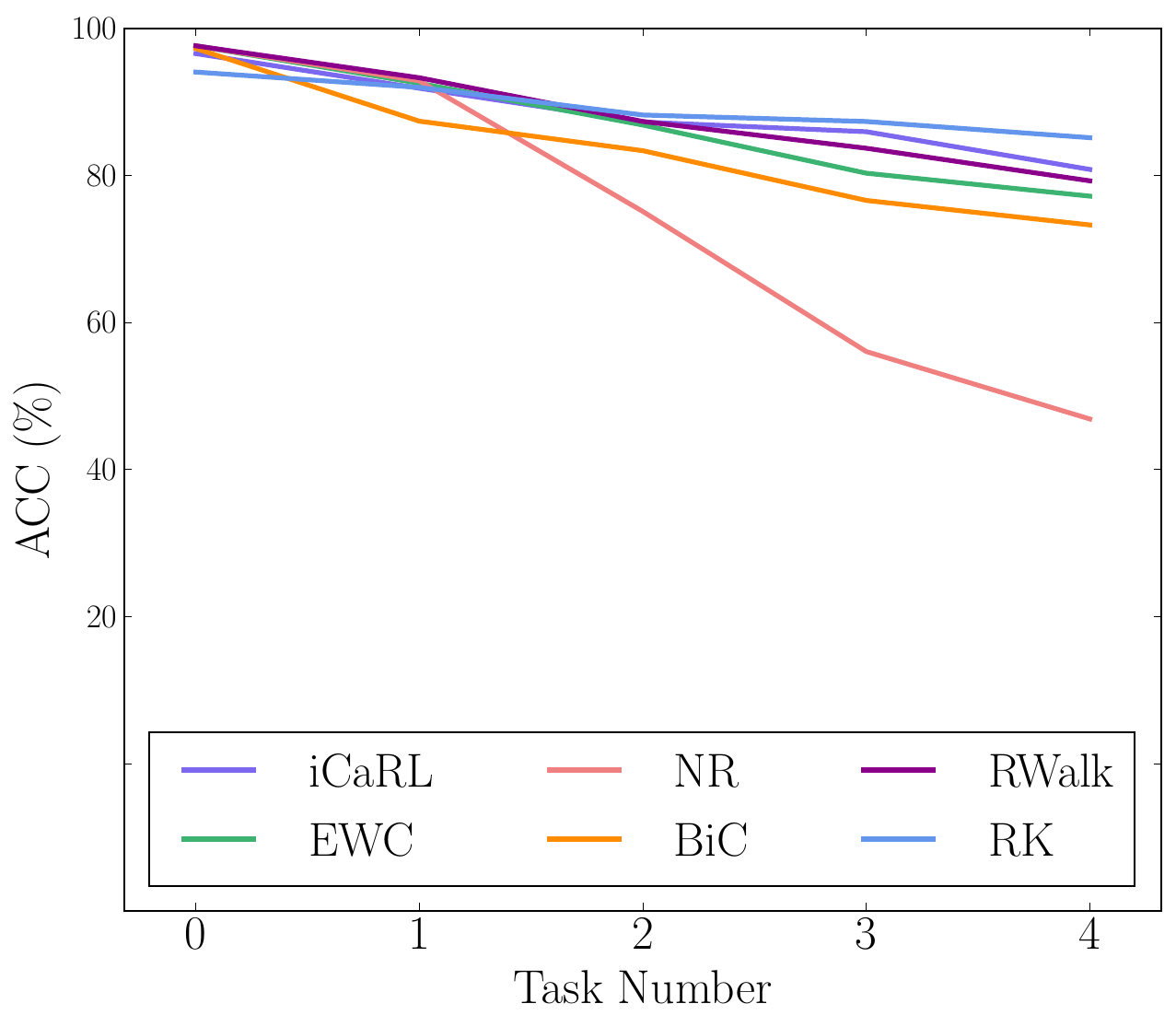}
  \centerline{(a)}
  \end{minipage}
\hfill
  \begin{minipage}[t]{0.49\linewidth}
  \centering
    \includegraphics[width=\linewidth]{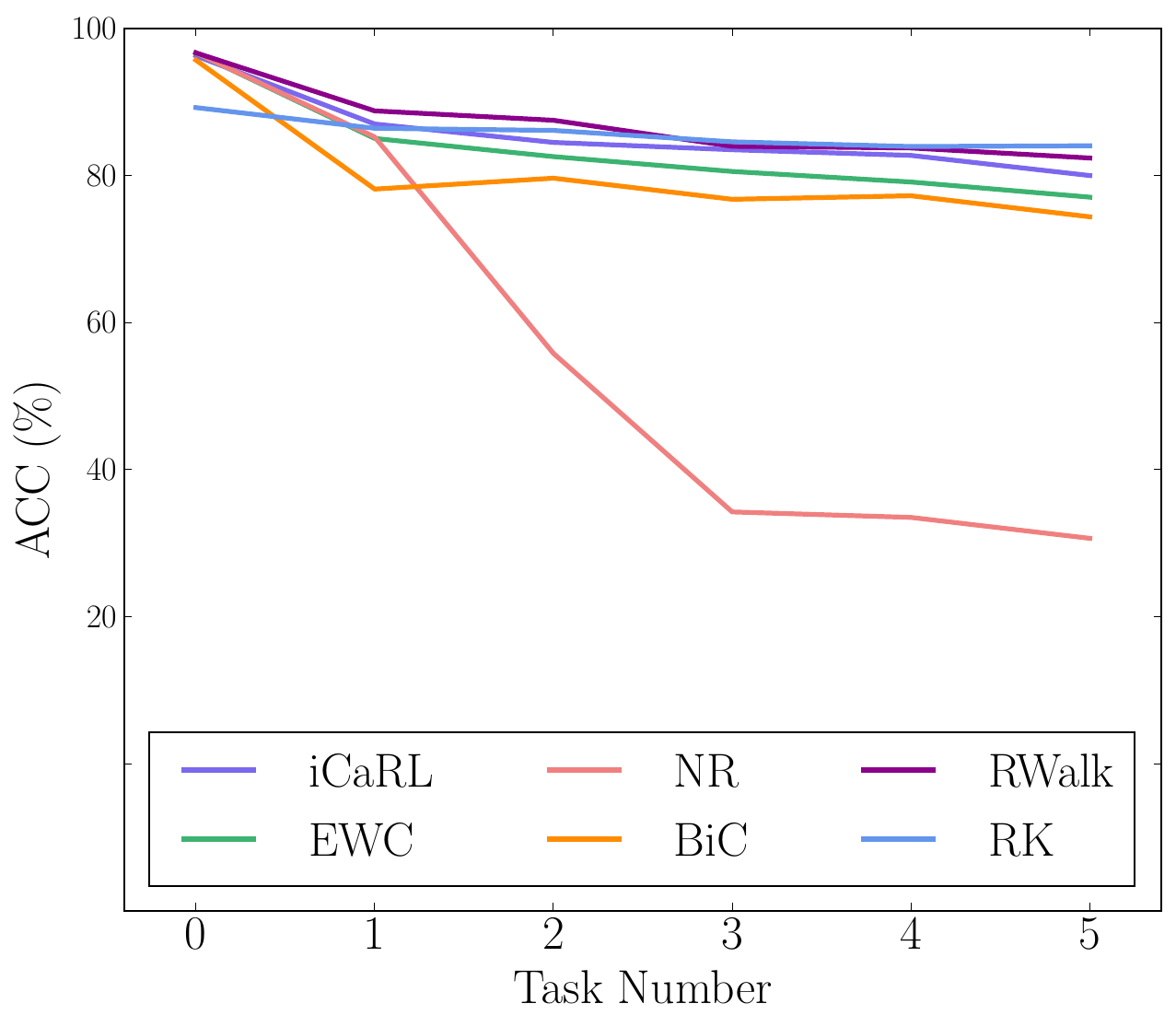}
  \centerline{(b)}    
  \end{minipage}
\hfill
  \begin{minipage}[t]{0.49\linewidth}
  \centering
  \includegraphics[width=\linewidth]{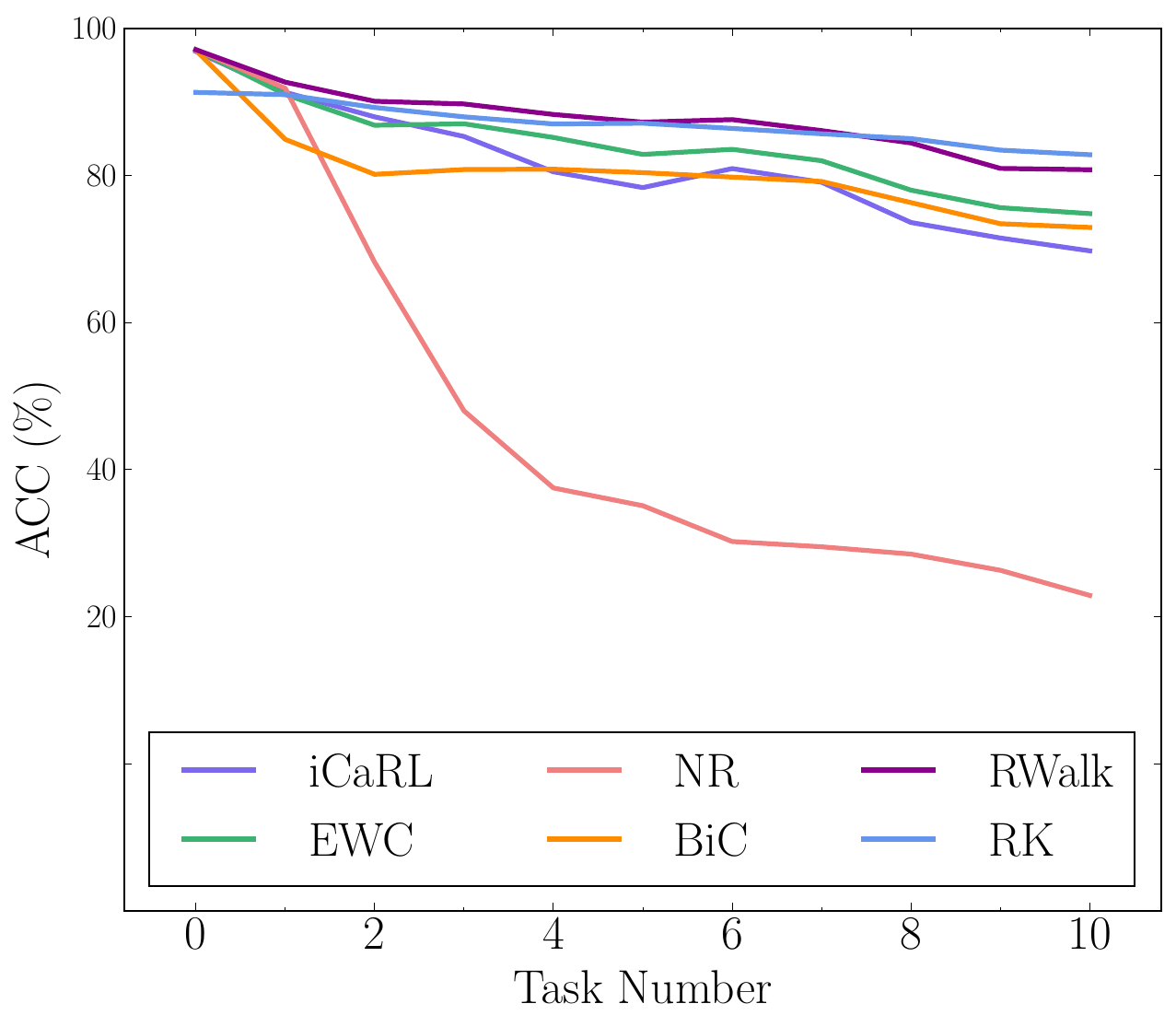}
  \centerline{(c)}
  \end{minipage}
\hfill
  \begin{minipage}[t]{0.49\linewidth}
  \centering
  \includegraphics[width=\linewidth]{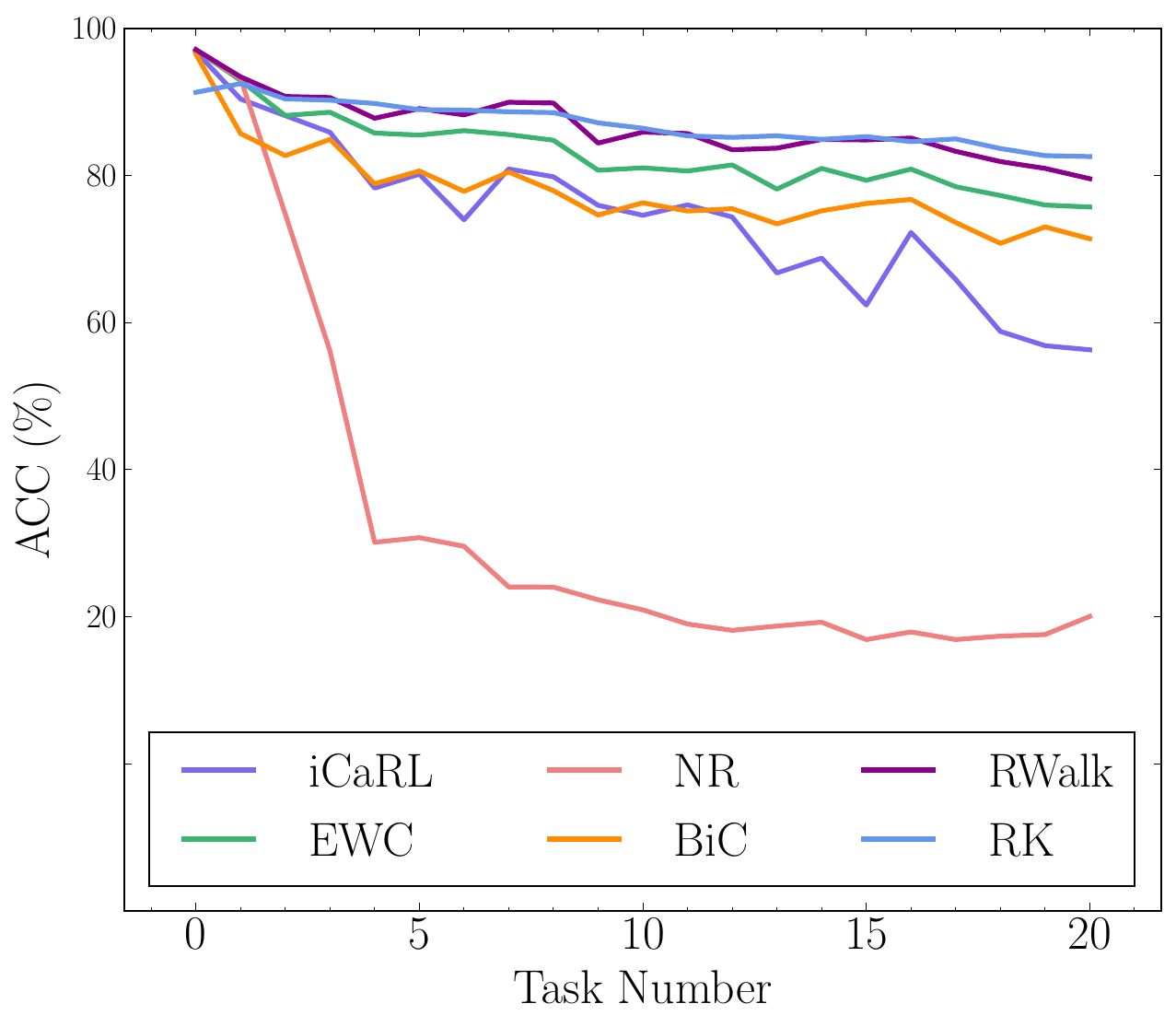}
  \centerline{(d)}
  \end{minipage}
  \caption{\textit{The ACC (\%) in a comparative study of increasing task numbers on the proposed RK approach and other competitive baselines. Figures from (a) to (d) represent the experiment with task numbers (= 20, 10, 5, 4).}}
  \label{fig:task}
\end{figure*}

\begin{figure}[!t]
   \centering
  \begin{minipage}[tbp]{0.49\linewidth}
  \centering
  \includegraphics[width=\linewidth]{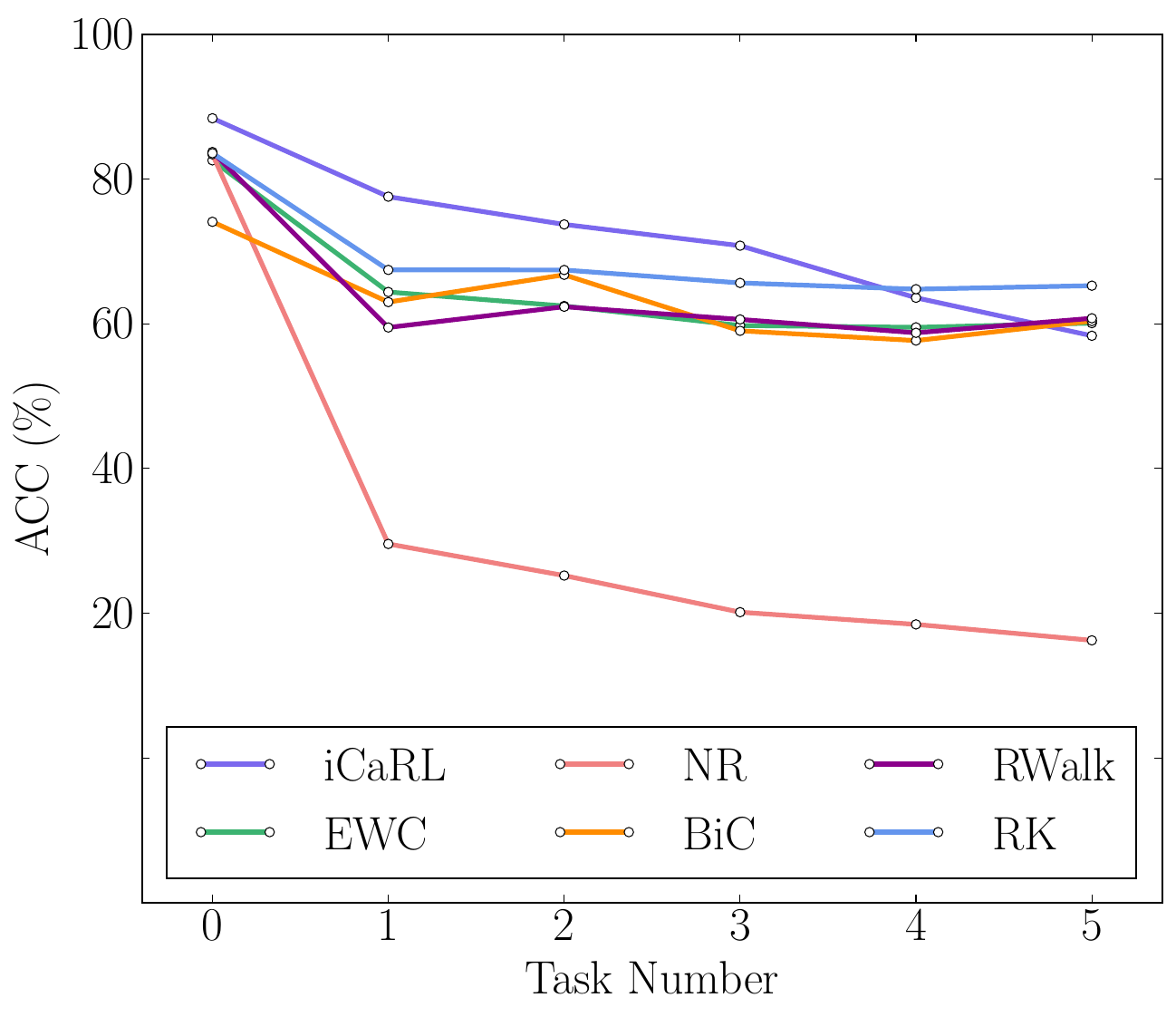}
  \centerline{(a) Memory size L = 300}
  \end{minipage}
\hfill
  \begin{minipage}[tbp]{0.49\linewidth}
  \centering
    \includegraphics[width=\linewidth]{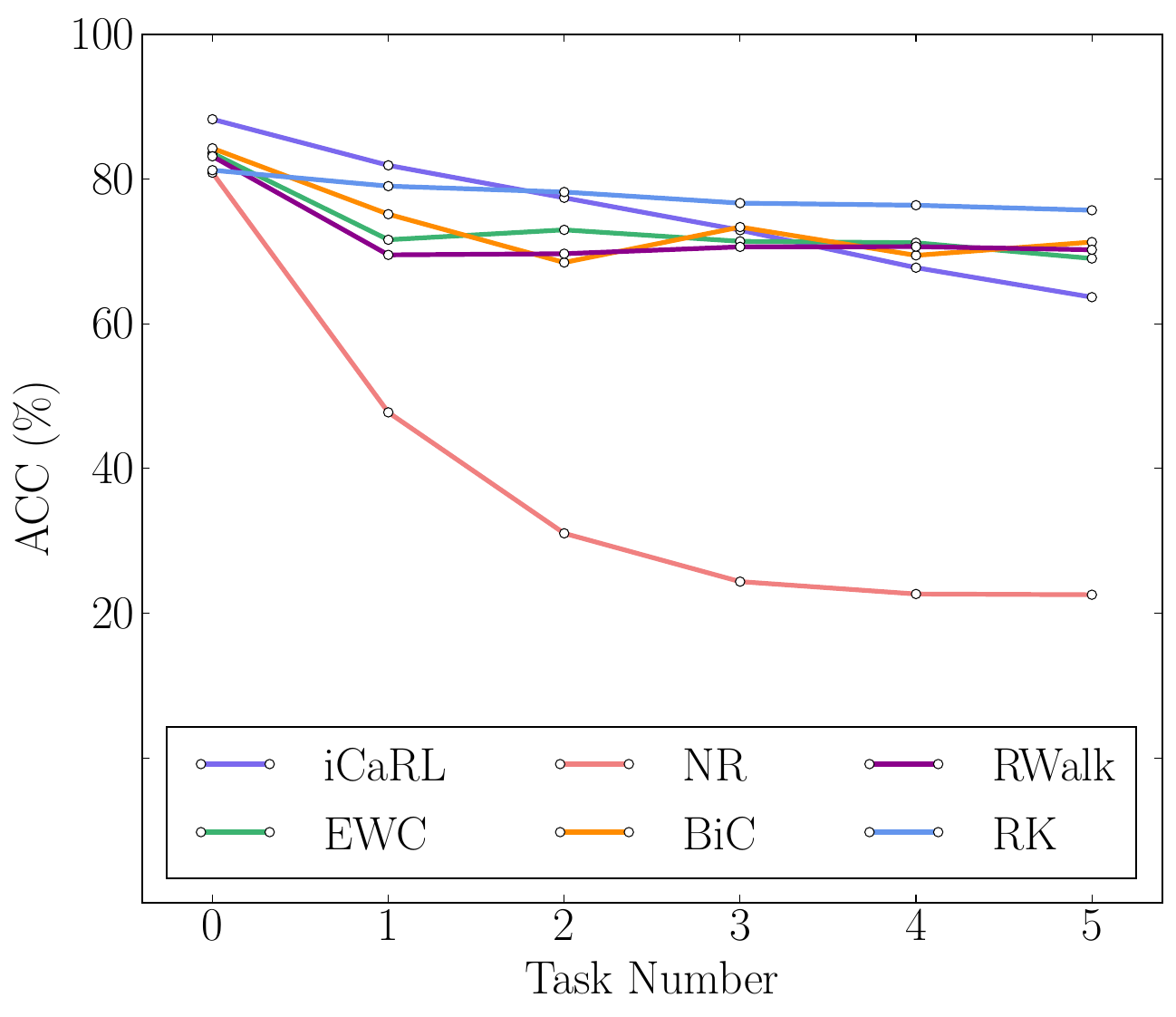}
  \centerline{(b) Memory size L = 500}    
  \end{minipage}
  \caption{\textit{The ACC (\%) in a comparative study of various memory size on the proposed RK approach and other competitive baselines.}}
  \label{fig:size}
\end{figure}

\subsection{Results}
\begin{table}[!t]
\Large
\centering
\caption{\textit{Average Accuracy (ACC) and Backward Transfer (BWT) in a comparative study of the proposed KD Loss. $L$ denotes the memory size for RK.}}
\label{tab:kd}

\begin{tabular}{l|c|c|c}
\hline
\hline
\textbf{Methods(L=500)} & \textbf{KD Loss} & \textbf{ACC($\uparrow$)} & \textbf{BWT($\uparrow$)} \\ \hline\hline
Rainbow Keywords & NO  & 0.779 & -0.033 \\ \hline
Rainbow Keywords & YES  & \textbf{0.779}         & \textbf{-0.015}                   \\ \hline\hline
\end{tabular}%

\end{table}

\subsubsection{Effect of the knowledge distillation loss}
We first analyse and summarize the performances with knowledge distillation loss. As shown in Table \ref{tab:kd}, we observe that the proposed knowledge distillation loss function achieves 54.5\% relative improvements in terms of BWT.

\subsubsection{Effect of various task numbers on RK approach}
We further analyse and summarize the performances of the proposed RK approach with increasing task numbers as shown in Figure \ref{fig:task}. The x-axis represents the task numbers (= 20, 10, 5, 4) and the y-axis is the evaluation metric of ACC. For example, when the task number is set to 20, we pre-train the TC-ResNet-8 with 21,000 audio clips including 10 unique keywords and the rest data is split into 20 tasks including 1 unique unseen keyword. The corresponding ACC is the accuracy of the testing set after each task is finished training. The memory size $L$ is 1500 for the proposed RK approach here. We observe that the proposed RK approach achieves the best ACC performances with increasing task numbers. Even though the difficulty of preserving prior knowledge is increased with the increasing task numbers, the proposed RK approach still can obtain over 85.0\% ACC, which is much better than other baselines.

\subsubsection{Effect of various memory size on RK approach}
We then report the effect of various memory size (L=300 or 500) on RK approach, as shown in Figure \ref{fig:size}. We constrain all methods under the same memory size as that of the RK approach. We observe that even all approaches perform much better with increasing memory size (i.e., more available training data), the proposed RK approach still outperforms other methods. Furthermore, even with limited memory size (L=300), the proposed RK approach can obtain over 65.0\% ACC performances, much better than other methods.
\begin{table}[!t]
\Large
\centering
\caption{\textit{Average Accuracy (ACC) and Backward Transfer (BWT) in a comparative study of the proposed data augmentation. ``NoAugment" means no data augmentation is applied in the experiment. $L$ denotes the memory size for RK.}}
\label{tab:da}

\begin{tabular}{l|c|c}
\hline\hline
\textbf{Methods(L=500)} & \textbf{ACC($\uparrow$)} & \textbf{BWT($\uparrow$)} \\ \hline\hline
NoAugment      & 0.779          & -0.033         \\ \hline
SpecAugment \cite{park2019specaugment}    & 0.783          & \textbf{-0.006 }    \\ \hline

Mixup \cite{zhang2017mixup} & \textbf{0.828} & -0.036  \\ \hline\hline
\end{tabular}%

\end{table}

\begin{table}[!t]
\Large
\centering
\caption{\textit{Accuracy and efficiency metrics in a comparative study of recent state-of-the-art training strategies. We adopt the TC-ResNet-8 model as testbed for all training strategies for fair comparison. Memory size $L$ is set to {[}500, 1500, 3000{]} in following experiments for RK.}}
\label{tab:fn}

\begin{tabular}{c|c|c|c|c}
\hline\hline
\textbf{Methods} & \textbf{ACC($\uparrow$)}   & \textbf{BWT($\uparrow$)}    & \textbf{Parameters} & \textbf{Memory} \\ \hline\hline
Fine-tune        & 0.262          & -0.372          & 64.48K              & 162.4M               \\ \hline
EWC              & 0.835          & -0.064         & 129.96K             & 162.4M               \\ \hline
PCL-KWS          & 0.836          & -0.041          & 406.9K              & 162.4M  
           \\ \hline
NR               & 0.560          & -0.163          & 64.48K              & 178.6M               \\ \hline
iCaRL            & 0.846          & -0.057          & 75.29K              & 178.6M               \\ \hline
BiC              & 0.793          & -0.085          & 64.48K                  & 178.6M               \\ \hline
RWalk            & 0.871          & -0.045          & 129.96K              & 178.6M    
             \\ \hline\hline
RK-500           & 0.828          & -0.036          & 129.96K              & \textbf{16.2M}       \\ \hline
RK-1500          & 0.887          & -0.012          & 129.96K              & 48.6M                \\ \hline
RK-3000          & \textbf{0.913} & \textbf{-0.012} & 129.96K              & 97.2M                \\ \hline\hline
Joint            & 0.940          & -               & 64.48K              & 1624.4M              \\ \hline
\end{tabular}%

\end{table}

\blfootnote{This research is supported by the National Research Foundation, Singapore under its AI Singapore Programme (AISG Award No: AISG-100E-2018-006).

The computational work for this article was partially performed on resources of the National Supercomputing Centre, Singapore (https://www.nscc.sg).}

\subsubsection{Effect of the data augmentation}
We also summarize the performances of the effects with two data augmentation methods on the proposed RK approach, as shown in Table \ref{tab:da}. We observe that all data augmentation methods improve the performances in terms of ACC. The best performances are achieved by the 'Mixup' data augmentation. We adopt the 'Mixup' data augmentation hereafter.

\subsubsection{Benchmark against other competitive methods}
Table \ref{tab:fn} summarizes the comparison between the proposed RK and other competitive methods in terms of Average Accuracy (ACC), Backward Transfer (BWT), Parameters and Memory. The task number is set to 5. We observe that the proposed RK achieves the best performance with memory size $L$ of 3000. Comparing with the best baselines: RWalk, the proposed RK-3000 achieves 4.2\% absolute improvements in terms of Average Accuracy with fewer memory, which is closer to the upper-bound performances. Furthermore, even with only 16.2M training data, our approach RK-500 has comparable performance to other baseline methods, which is effective on edge devices. 

\section{Summary}
In this paper, we propose a novel diversity-aware class incremental learning method named Rainbow Keywords (RK) approach to avoid catastrophic forgetting with less memory. Experimental results show that the proposed RK approach achieves 4.2\% absolute improvement in terms of average accuracy over the best baseline. Ablation study also indicates that the proposed data augmentation and knowledge distillation loss are quite effective on edge devices. For future work, we plan to apply and adapt our approach to multilingual keyword search systems \cite{nancy1,nancy2,nancy3}.
\newpage

%% file: Chapter4/Chapter4.tex

\chapter{Continual learning for environmental sound classification}
In this chapter, we investigate the replay-based CL (RCL) methods for on-device environmental sound classification. We first study the performance of existing memory update algorithm (MUA) methods such as \textit{Reservoir} \cite{vitter1985random}, \textit{Prototype} \cite{rebuffi2017icarl} and \textit{Uncertainty} \cite{bang2021rainbow} (as described in Section \ref{sec:mua}) on RCL for on-device environmental sound classification. \\

\noindent We empirically demonstrate that \textit{Uncertainty} \cite{bang2021rainbow} method performs best in our scenario. Furthermore, we propose \textit{Uncertainty++}, a simple yet efficient MUA method based on \textit{Uncertainty} method. Different to the \textit{Uncertainty} method, our proposed \textit{Uncertainty++} introduces the perturbations to the embedding layer of the classifier. As a result, the computation cost (e.g., running memory and time) can be significantly reduced when measuring the data uncertainty. We evaluate the performance of our method on the DCASE 2019 Task1 \cite{tau2019} and the ESC-50 \cite{piczak2015esc} datasets with on-device model BC-ResNet-Mod ($\sim$86k parameters) \cite{kim2021broadcasted,Kim2021b}. Experimental results show that \textit{uncertainty++} outperforms the existing MUA methods on classification accuracy, indicating its potential in real-world on-device audio applications. Our proposed method is model-independent and simple to apply. Our code is made available at the GitHub\footnote{\url{https://github.com/swagshaw/ASC-CL}}.

\section{Related work}
Recently, replay-based CL methods have shown promising results outperforming regularization-based methods in audio tasks such as keywords spotting \cite{huang2022progressive, xiao2022rainbow} and sound event detection \cite{wang2019continual}. However, CL in on-device applications, such as on-device environmental sound classification, has received less attention in the literature, which is the focus in this paper. The on-device scenarios are often associated with restrictions in storage and memory space \cite{singh2022passive}, which can pose challenges to replay-based CL which relied on external memory to restore historical data. As a result, the sound classification models that can be operated on the device may be limited in their capacities, thus prone to forgetting old knowledge when continuously learning new sound classes.
\section{Method}
\label{sec:method}
This section first describes replay-based continual learning and four memory update algorithms, and then introduces the proposed \textit{uncertainty++} algorithm.
\subsection{Replay-based continual learning}
\label{sec:mua}
Following the continual learning setting \cite{awasthi2019continual,zenke2017continual,wang2019continual} of environmental sound classification, we assume that the model M should identify all classes in a series of tasks $T=\{\tau_0,\dots,\tau_t\}$ without catastrophic forgetting. For each task $\tau\in T$, we have input pairs $(x,y)$ and classes $C$, where $x$ denotes audio waveforms and $y$ are classes $c\in C$. We aim to minimize a cross-entropy loss of all classes $C$ present in the current task $\tau$ formulated as:\\
\begin{equation}
L_{CE}(\tau) = \sum\limits_{c\in C} y_c log \frac{exp{(M(x)}_c)}{\sum\limits_{c\in C} exp{(M(x)}_c)},
\label{eq4}
\end{equation}

\noindent Where $M(x)$ denotes the output of the model $M$ for input $x$.\\

\noindent The parameters learned from the previous task are potentially overwritten after learning the new class, also known as catastrophic forgetting. To mitigate this issue, we introduce replay-based methods. The replay-based methods utilize a region of the memory which is called `replay buffer' to temporarily store the historical training samples to maintain the performance.\\

\noindent Re-training sound classification models with the mixture of the whole historical and new data is resource- and time-consuming in real-world on-device scenarios. To mitigate this issue, the replay-based methods access only a subset of the historical data to save the storage space. In this case, how to select the part of samples to the replay buffer by the memory update algorithm is the key.\\

\noindent Specifically, in the training of task $\tau_t$, the replay buffer stores the selected training samples from the previous $t-1$ learned task(s) $\{\tau_0,\tau_1,\dots,\tau_{t-1}\}$, and builds the training data buffer $\hat D_t$ for task $\tau_t$ formulated as:\\
\begin{equation}
\label{eq2}
\hat D_t = g(\hat D_{t-1})\cup D_t,
\end{equation}
\noindent where $g$ is the memory update algorithm \cite{xiao2022rainbow}, $\hat D_{t-1}$ is the training data buffer for task $\tau_{t-1}$, and $D_t$ is the incoming data for the new task. 

\subsubsection{Memory update algorithm (MUA)}
We introduce four memory update algorithms in the literature. Generally, we assume that the memory update should select $L$ samples from the training data $\hat D_{t-1}$ of the previous task $\tau_{t-1}$ for the training of the task $\tau_t$. \\

\noindent \textbf{\textit{Random}} \cite{hsu2018nr} memory update algorithm selects $L$ new samples $\{(x_1,y_1),(x_2,y_2),\dots,(x_L,y_L)\}$ for the next task randomly from the candidates $\hat D_{t-1}$ into replay buffer.\\

\noindent \textbf{\textit{Reservoir}} \cite{vitter1985random} memory update algorithm conducts uniform sampling from $\hat D_{t-1}$. Specifically, the reservoir algorithm initializes the replay buffer indexed from 1 to $L$, containing the first $L$ items $\{(x_1,y_1),(x_2,y_2),\dots,(x_L,y_L)\}$ of the candidates. When updating replay buffer from the candidates, for each sample, the reservoir algorithm generates a random number $m$ uniformly in $\{1,\dots, len(\hat D_{t-1})\}$. If $m \in \{1,\dots,L\}$, then the sample with the index $m$ in the replay buffer is replaced with the sample $\hat D_{t-1}[m]$.\\

\noindent \textbf{\textit{Prototype}} \cite{rebuffi2017icarl} 
 memory update algorithm selects the samples from $\hat D_{t-1}$ where the embedding of the classifier is close to the embedding mean of its own class. Specifically, the algorithm first groups the $\hat D_{t-1}$ into subsets as $D_c, c=1\dots N^t$ by unique classes, where $N^t$ denotes the total numbers of unique classes in the $\hat D_{t-1}$ set. Then the algorithm uses the current model to extract the embedding of the candidates for each $D_c$ and calculates the class mean by the embedding as the average feature vector. For each class, the algorithm selects the samples of the candidates so that the average feature vector over the replay buffer provides best approximate to the average feature vector over all the samples of the corresponding class.\\

\noindent\textbf{\textit{Uncertainty}} \cite{bang2021rainbow} memory update algorithm selects the sample by the uncertainty of the sample through the inference by the classification model. Specifically, the first step groups the $\hat D_{t-1}$ in the same way as the \textit{prototype} algorithm introduced above. The second step estimates the uncertainty of each sample $x$ in $D_c$. Predictive likelihood captures how well a model fits the data, with larger values indicating better model fit. Uncertainty score can be determined from predictive likelihood \cite{gal2016dropout}. Following the derivation from \cite{gal2016dropout}, the predictive likelihood of a sample given by the model can be approximated by the Monte-Carlo (MC) integration \cite{kloek1978bayesian} method with the model outputs of perturbed samples \cite{xiao2022rainbow}, which is defined as follows:\\
\begin{equation}
\label{eq5}
    P(y = c \mid x) = \int p(y=c \mid \hat{x})p(\hat{x}\mid x) d\hat{x},
\end{equation}
\noindent where $x$, $\hat{x}$, $y$ denote an audio utterance of one class, the perturbed samples of $x$, and the label of $x$. Therefore, the uncertainty of the audio utterance $x$ is formulated as $u(x)$: \\
\begin{equation}
\label{ux1}
    u(x) \approx 1- \frac{1}{K} \sum\limits^K_{k=1} P(y = c\mid \hat{x}_k),
\end{equation}
\noindent where $K$ presents the number of the perturbations generated by perturb methods such as Audio Shift \cite{ko2015audio}, Audio PitchShift \cite{ko2015audio} and Audio Colored Noise \cite{audiomentation,kamath2002multi}. A larger $u(x)$ indicates a smaller confidence of the model in predicting the perturbed samples.\\
\begin{figure}[]
  \centering
  \includegraphics[width=\linewidth]{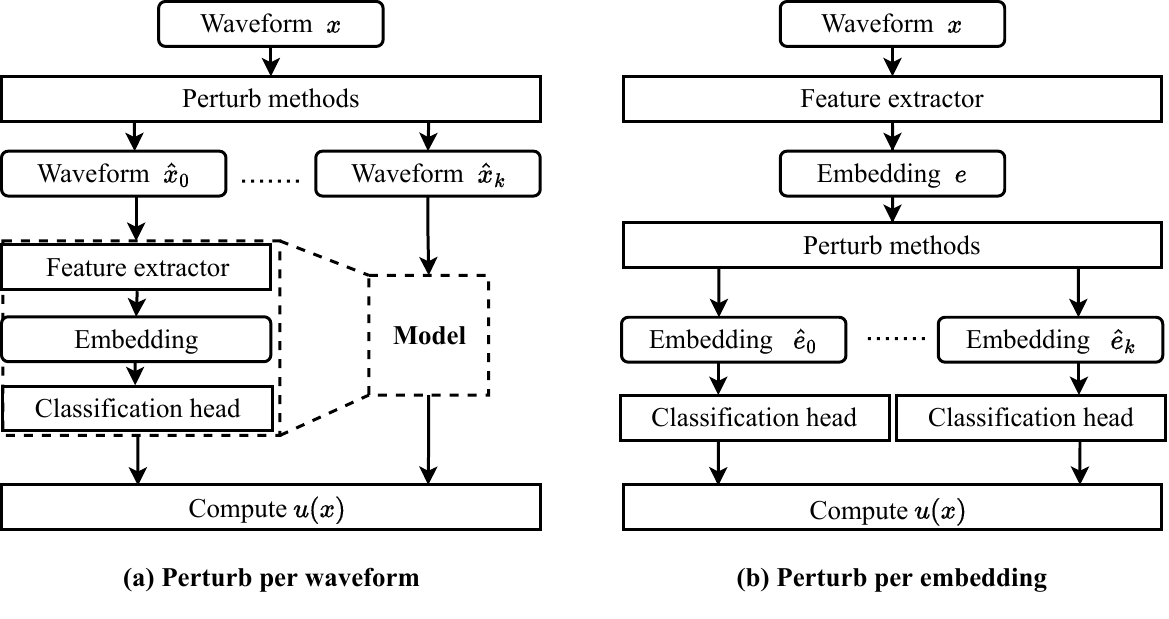}
  \caption{\textit{Block diagram of the native uncertainty approach and our proposed approach. Specifically, the naive approach adds perturbations to $x$ by waveform and generates multiple waveform as $\hat x$. Our approach inputs the embedding $e$ and generates perturbed embedding $\hat e$ which means we only save the embedding. The output of the backbone of the model is calculated only once.  ``Compute $u(x)$" is to compute $u(x)$ by Eq. (\ref{eq5}). The $K$ refers to the number of the perturbations generated by perturb methods.}}
  \label{fig:workflow1}
\end{figure}

\noindent The third step selects $L$ examples from $D_c$ through descending the uncertainty $u(x)$ with the step size of $len(D_c)*C/L$, where $L$ is the size of the replay buffer.\\

\noindent Previous research \cite{xiao2022rainbow} demonstrated that the uncertainty memory update algorithm performs better than the other three algorithms on speech tasks such as keyword spotting. However, the computation cost of \textit{Uncertainty} increases linearly with the number of perturbation operations.\\

\subsection{Proposed MUA method (\textit{Uncertainty++)}}
 As illustrated in Figure \ref{fig:workflow1}, the native uncertainty memory update algorithm requires to employ perturbation methods offline for the waveform of each sample to generate the perturbed samples first. In our proposed method, noisy perturbations are added to the pre-classifier embedding of the sample, and not to the waveform, so the output of the backbone of the model is calculated only once. Specifically, we propose a vector-wise perturbation method that adds noise with different intensities according to the variance of classifier's embedding. We denote the perturbed version of the classifier's embedding $e$ as $\hat e$, which is computed as follows:\\
 \begin{equation}
\hat e = e + U(-\frac{\lambda}{2},\frac{\lambda}{2})*std(e),
\end{equation}
\noindent where $std(\cdot)$ stands for standard deviation, the function $U(a,b)$ represents the noise distributed uniformly from $a$ to $b$, $U(a,b)$ is a vector with the same shape as $e$, and $\lambda$ is a hyperparameter that controls the relative noise intensity. \\

\noindent By the vector-wise perturbation method, we generate the perturbed embedding $\hat e$ of the embedding $e$. Finally, we input $\hat e$ to the final classification layer of the model and output $P(y=c\mid\hat{e})$ which is used to compute the uncertainty as in Eq. (\ref{eq5}). After the uncertainty is estimated, we select examples for replay as native approach. This method saves time by calculating the output of the backbone of the model only once. We also save the memory usage by replacing the extra perturbed raw data with the classifier's embedding which is of much smaller size as compared with the raw data.\\
\section{Experiments and Results}
\label{sec:exp}
\subsection{Environmental sound classification model}
For the on-device environmental sound classification model, we use BC-ResNet-Mod \cite{Kim2021b} which is an adaptation of the BC-ResNet \cite{kim2021broadcasted} that achieves improved results on acoustic scene classification. The BC-ResNet paradigm works via repeatedly extracting spectral and then temporal features in series. Because these spectral features are of a lower dimension than the input, this model has fewer parameters than one that processes the waveform directly. Feature extraction is channel-wise, and both parameter reductions have negligible impact on performance \cite{kim2021broadcasted}. For our experiments, we use BC-ResNet-Mod-4, which increases the input channel dimension to 80 before extracting spectral and temporal features.
\subsection{Datasets}
\textbf{ESC-50} consists of \num{2000} five-second environmental audio recordings \cite{piczak2015esc}. Data are balanced between \num{50} classes, with \num{40} examples per class, covering animal sounds, natural soundscapes, human sounds (non-speech), and ambient noises. The dataset has been prearranged into five folds for cross-validation. \\

\noindent \textbf{DCASE 2019 Task 1} is an acoustic scene classification task, with a development set \cite{tau2019} consisting of \num{10}-second audio segments from \num{10} acoustic scenes: airport, indoor shopping mall, metro station, pedestrian street, public square, the street with a medium level of traffic, traveling by tram, traveling by bus, traveling by an underground metro and urban park. In the development set, there are \num{9185} and \num{4185} audio clips for training and validation, respectively.\\
\begin{table}[t]
\centering
\caption{\textit{Accuracy (ACC) and Backward Transfer (BWT) in a comparative study of the proposed memory update algorithm.}}
\Large
\begin{tabular}{@{}ccccc@{}}
\toprule
\multirow{2}{*}{\textbf{Method}} & \multicolumn{2}{c}{\textbf{DCASE 2019 Task 1}} & \multicolumn{2}{c}{\textbf{ESC-50}} \\ \cmidrule(l){2-5} 
                        & \textbf{ACC $\uparrow$}              & \textbf{BWT} $\uparrow$             & \textbf{ACC} $\uparrow$          & \textbf{BWT} $\uparrow$         \\ \cmidrule{1-5}
\textit{Finetune}             & 0.205          & -0.276          & 0.181          & -0.307          \\ \hline
\textit{Random}               & 0.473          & -0.115          & 0.225          & -0.231          \\ \hline
\textit{Reservoir}            & 0.568          & -0.096          & 0.430          & -0.121          \\ \hline
\textit{Prototype}            & 0.559          & -0.089          & 0.482          & \textbf{-0.104} \\ \hline
\textit{Uncertainty} & 0.578 & \textbf{-0.079} & 0.477 & -0.111          \\  \hline
\textit{Uncertainty++} & \textbf{0.581} & \textbf{-0.079} & \textbf{0.500} & -0.121          \\ 
\hline
\end{tabular}

\end{table}

\subsection{Experimental setup}
\textbf{Task setting} To evaluate the performance of the proposed approach, we split the data into five tasks. Each task includes \num{2} new unique classes in DCASE 19 Task 1 and \num{10} new unique classes in ESC-50, which is unseen in previous tasks. To simulate the condition of edge devices, we set the buffer size $L$ of examples as \num{500}, \num{100} samples in DCASE 19 Task 1 and ESC-50 due to the memory limitation. \\

\noindent \textbf{Implementation details} The original audio clip is converted to \num{64}-dimensional log Mel-spectrogram by using the short-time Fourier transform with a frame size of \num{1024} samples, a hop size of \num{320} samples, and a Hanning window. The classification network is optimized by the Adam \cite{kingma2014adam} algorithm with the learning rate \num{1e-3}. The batch size is set to \num{32} and the number of epochs is \num{50}. \\
\subsection{Evaluation metrics}
We report performances in terms of the accuracy and forgetting metric. Specifically, the \textit{Accuracy} (ACC) reports an accuracy averaged on learned classes after the entire training ends. The \textit{Backward Transfer} (BWT) \cite{lopez2017gradient} evaluates accuracy changes on all previous tasks after learning a new task, indicating the forgetting degree.  For measuring BWT, we first construct the matrix $R \in \mathbb{R}^{T\times T}$, where $R_{i,j}$ is the test classification accuracy of the model on task $\tau_j$ after observing the last sample from task $\tau_i$. After the model finished learning about each task $\tau_i$, we evaluate its BWT on all $T$ tasks, which is formulated as:\\
\begin{equation}
\begin{aligned}
BWT = \frac{1}{T-1} \sum\limits^{T-1}_{i=1} R_{T,i} - R_{i,i}.
\end{aligned}
\end{equation}

\noindent There exists negative BWT when learning about some task decreases the performance on some preceding task. A smaller value of BWT indicates a higher catastrophic forgetting.\\
\subsection{Reference baselines}
We built five baselines for comparisons. The \textit{Finetune} training strategy adapts the BC-ResNet-Mod model for each new task without any continual learning strategies, as the lower-bound baseline. The four prior memory update algorithms of replay-based continual learning (i.e., \textit{Random}, \textit{Reservoir}, \textit{Prototype}, \textit{Uncertainty}) are introduced in Section \ref{sec:mua}. Specifically, at the perturbation stage of the uncertainty, we use two perturbation methods, namely, `\textit{uncertainty-shift}', which includes Audio Shift and Audio PitchShift, and  `\textit{uncertainty-noise}' which refers to the Audio Colored Noise perturbation method.\\

\begin{table}[t]
\centering
\caption{\textit{Accuracy (ACC) and Backward Transfer (BWT) in a comparative study of the proposed perturbation method. The $K$ refers to the number of the perturbations generated by perturbation methods.}}
\Large
\begin{tabular}{@{}lccccc@{}}
\toprule
\multicolumn{1}{c}{\multirow{2}{*}{\textbf{Method}}}  &
\multicolumn{1}{c}{\multirow{2}{*}{\textbf{K}}} &
\multicolumn{2}{c}{\textbf{DCASE 2019 Task 1}} & \multicolumn{2}{c}{\textbf{ESC-50}} \\ \cmidrule(l){3-6} 
                     &   & \textbf{ACC $\uparrow$}              & \textbf{BWT} $\uparrow$             & \textbf{ACC} $\uparrow$          & \textbf{BWT} $\uparrow$         \\ \cmidrule{1-6}
\multicolumn{1}{c}{} & 2                 & 0.557                 & -0.101                & 0.461
             & \textbf{-0.111}
            \\
\multicolumn{1}{c}{\textit{Uncertainty-Shift}} & 4                 & 0.575                 & -0.103                & 0.476
             & -0.118
             \\
\multicolumn{1}{c}{} & 6                 & 0.567                 & -0.079                & 0.477
             & -0.118
            \\ \hline
\multicolumn{1}{c}{} & 2                 & 0.560                 & -0.100                & 0.465
             & -0.118            \\
\multicolumn{1}{c}{\textit{Uncertainty-Noise}} & 4                 & 0.535                 & -0.104                & 0.473
             & -0.118
            \\
\multicolumn{1}{c}{} & 6                 & 0.578                 & -0.079                & 0.458
             & -0.120
            \\ \hline
\multicolumn{1}{c}{} & 2             & 0.571                 & -0.102                & \textbf{0.500}
              & -0.121
            \\
\multicolumn{1}{c}{\textit{Uncertainty++}} & 4            & 0.548                 & -0.103                & 0.481
             & -0.114
            \\
\multicolumn{1}{c}{} & 6             & \textbf{0.581}                 & \textbf{-0.079}                 & 0.484
              & -0.119            \\ \bottomrule
\end{tabular}

\end{table}

\subsection{Results}
\label{sec:res}
\subsubsection{Experiments on MUA methods}
Table 1 presents the results on DCASE 2019 Task 1 and ESC-50 test set in terms of ACC and BWT. We compare the proposed \textit{Uncertainty++} MUA method with five baselines. We observe that the \textit{uncertainty} MUA method achieves better performance than the five baselines. Comparing with the best baseline \textit{uncertainty}, we observe that the proposed \textit{uncertainty++} method obtains 58.1\% on classification accuracy which outperforms the existing MUA methods. In addition, we observe that the \textit{Finetune} method achieves the worst ACC and BWT performance compared with other baselines, which indicates the issue of catastrophic forgetting.\\

\noindent We further analyze and summarize the performances of the proposed \textit{uncertainty++} method compared with the \textit{uncertainty} MUA method with different numbers of the perturbation methods in terms of ACC and BWT as shown in Table 2. The $K$ refers to the number of the perturbations generated by perturb methods. Even with only two perturbation methods, our proposed method still outperforms other two baselines. We also observe that our method under two perturbations obtains the best performance on the ESC-50 test set. Such performance might be due to the small size of the ESC-50, therefore it is more sensitive to perturbations.\\

\subsubsection{Comparative experiments on computation time for \textit{Uncertainty} and \textit{Uncertainty++}}
We further report the Average Time for the proposed method when there is an increasing number of perturbations.
The Average Time measures a relative time increase compared to training time in each task. As shown in Table 3, even with $6$ perturbations, the Average Time of the \textit{uncertainty++} is still less than \num{60}s. This can be explained by the fact that our proposed method can limit the growth of the additional training time. We also observe that our proposed method outperforms other baselines in any number of perturbations, which indicates our proposed method is computationally more efficient. In addition, the average time of \textit{uncertainty-shift} is much longer than others. Because the Audio Shift and Audio PitchShift perturbations takes more time than simply adding noise.

\begin{table}[]
\centering
\caption{\textit{Average Time (s) in a comparative study of the proposed \textit{uncertainty++} method. The $K$ refers to the number of the perturbations generated by perturbation methods.}}
\Large
\begin{tabular}{@{}lccc}
\toprule
\multicolumn{1}{c}{\textbf{Method}}  & \multicolumn{1}{c}{\textbf{K}} & \multicolumn{1}{c}{\textbf{Average Time (s) $\downarrow$}}         \\ \cmidrule{1-3}
\multicolumn{1}{c}{} & 2                 & 1221.7                
            \\
\multicolumn{1}{c}{\textit{Uncertainty-Shift}} & 4                 & 2205.1                
             \\
\multicolumn{1}{c}{} & 6                 & 2926.1                
            \\ \hline
\multicolumn{1}{c}{} & 2                 & 246.2                           \\
\multicolumn{1}{c}{\textit{Uncertainty-Noise}} & 4                 & 390.8               
            \\
\multicolumn{1}{c}{} & 6                 & 506.3                
            \\ \hline
\multicolumn{1}{c}{} & 2 & 44.0            
            \\
\multicolumn{1}{c}{\textit{Uncertainty++}} & 4            & 48.5               
            \\
\multicolumn{1}{c}{} & 6             & 55.1                \\ \bottomrule
\end{tabular}

\label{tab:time}
\end{table}
\section{Summary}
In this chapter, we have presented \textit{uncertainty++}, an efficient replay-based continual learning method for on-device environmental sound classification. Our method selects the historical data for the training by measuring the per-sample classification uncertainty on the embedding layer of the classifier. Experimental results on the DCASE 2019 Task 1 and ESC-50 datasets show that our proposed method outperforms the baseline continual learning methods on classification accuracy and computational efficiency. In future work, we plan to apply and adapt our approach to other on-device audio classification tasks such as audio tagging and sound event detection.
\label{sec:conclusion}
\newpage

%% file: Chapter5/Chapter5.tex

\chapter{Conclusion and Recommendations}

\section{Conclusion}

Current deep-learning-based audio classification systems are usually trained with limited classes in the compact model for lower computation and smaller footprint. Therefore, the performance of the model trained by the source-domain data may degrade significantly
when confronted with unseen classes of the target-domain at run-time. The naive approach of fine-tuning, so fruitfully applied to domain transfer problems, suffers from the lack of data from previous tasks and the resulting classifier is unable to classify data from them. This drastic drop in performance on previously learned tasks is a phenomenon known as catastrophic forgetting \cite{mccloskey1989catastrophic}.  Continual learning aims to prevent catastrophic forgetting. In this thesis, we first investigate the recent development of the two of audio classification tasks -- keyword spotting and environmental sound classification. Then we review the several categories of continual learning approaches. For the continual learning of keyword spotting, we propose a novel diversity-aware class incremental learning method named Rainbow Keywords (RK) approach to avoid catastrophic forgetting with less memory. Experimental results show that the proposed RK approach achieves 4.2\% absolute improvement in terms of average accuracy over the best baseline. Ablation study also indicates that the proposed data augmentation and knowledge distillation loss are quite effective on edge devices. On the other hand, for environmental sound classification, we present uncertainty++, an efficient replay-based continual learning method for on-device environmental sound classification. Our method selects the historical data for the training by measuring the per-sample classification uncertainty on the embedding layer of the classifier. Experimental results on the DCASE 2019 Task 1 and ESC-50 datasets show that our proposed method outperforms the baseline continual learning methods on classification accuracy and computational efficiency. \\

\section{Future Work}
\subsection{Continual learning on different SNR conditions}
After this work, I propose to investigate the effect of continual learning on
different SNR conditions to make the KWS system adaptive and continual
in both clean and noisy environments. Crucially, we always train on the full set of
different SNR levels, only choosing between orderings. To execute, we divide the training
process into five progressively harder steps. At the start, we conditioned the model on
clean samples without noise. And using continual learning methods to keep the model
maintain the previous knowledge. As the result, the model can not only contain better
performance on the clean condition but also perform better in the noisy environment.

\subsection{Continual learning on multilingual}
\cite{2021Few} introduce a few-shot transfer learning method for keyword spotting in any language. They train an embedding model on keyword classification using Common Voice’s \cite{ardila2020common} multilingual crowd-sourced speech dataset, by applying forced alignment \cite{mcauliffe2017Forced} to automatically extract 760 frequent words across nine languages. They then finetune this embedding model to classify a target keyword with just five sample utterances, even if the model has never seen the target language before.
Across 440
keywords in 22 languages, they achieve an average streaming
keyword spotting accuracy of 87.4\% with a false acceptance
rate of 4.3\%, and observe promising initial results on keyword
search. \\

\noindent Recently, there are other more cross-lingual learning methods which is suitable for multilingual KWS. Cross-lingual learning aims to build models which leverage data from other languages to improve
performance. XLS-R \cite{babu2021xls} is a large-scale model for cross-lingual speech representation
learning based on wav2vec 2.0. They train models with up to 2B parameters on
nearly half a million hours of publicly available speech audio in 128 languages, an order of magnitude more public data than the largest known prior work. Moreover, they show that with sufficient model size,
cross-lingual pretraining can perform as well as English-only pretraining when
translating English speech into other languages, a setting which favors monolingual
pretraining. XLS-R can help to improve speech processing tasks for many
more languages of the world including KWS.\\

\noindent I propose to combine the cross-lingual pretraining model XLS-R and the wav2kws \cite{seo2021wav2kws} for the multilingual KWS task. Also, I propose to explore the continual learning method for the multilingual KWS. For example, solving the forgetting issue when learning a series of languages.

\newpage

%% file: Appendix/appendix.tex

\chapter*{Appendix A}
\addcontentsline{toc}{chapter}{Appendix A}

\begin{figure}[htbp]
  \centering
  \includegraphics[width=\linewidth]{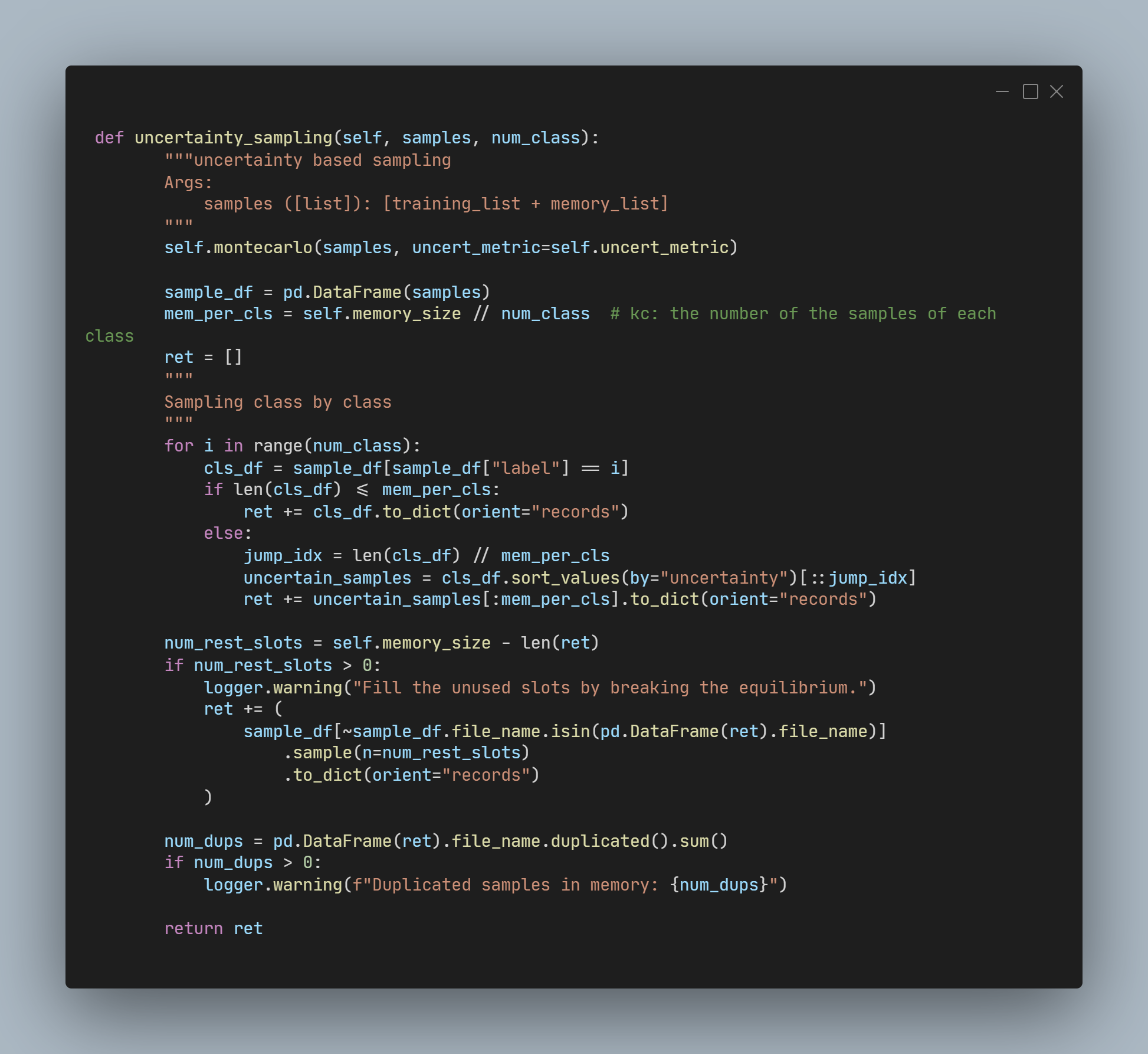}
  \caption{\textit{The python code of the uncertainty sampler.}}
  \label{fig:sampler}
\end{figure}


\newpage

%% file: Ref/References.bib
@article{mesaros2021sound,
  title={Sound event detection: A tutorial},
  author={Mesaros, Annamaria and Heittola, Toni and Virtanen, Tuomas and Plumbley, Mark D},
  journal={IEEE Signal Processing Magazine},
  volume={38},
  number={5},
  pages={67--83},
  year={2021},
  publisher={IEEE}
}

@article{liu2022simple,
  title={Simple Pooling Front-ends For Efficient Audio Classification},
  author={Liu, Xubo and Liu, Haohe and Kong, Qiuqiang and Mei, Xinhao and Plumbley, Mark D and Wang, Wenwu},
  journal={arXiv preprint:2210.00943},
  year={2022}
}

@article{lopez2021deep,
  title={Deep spoken keyword spotting: An overview},
  author={L{\'o}pez-Espejo, Iv{\'a}n and Tan, Zheng-Hua and Hansen, John and Jensen, Jesper},
  journal={IEEE Access},
  year={2021},
  publisher={IEEE}
}

@inproceedings{radhakrishnan2005audio,
  title={Audio analysis for surveillance applications},
  author={Radhakrishnan, Regunathan and Divakaran, Ajay and Smaragdis, A},
  booktitle={Proc. IEEE Workshop on Applications of Signal Processing to Audio and Acoustics},
  pages={158--161},
  year={2005},
  organization={IEEE}
}

@inproceedings{peng2009healthcare,
  title={Healthcare audio event classification using hidden Markov models and hierarchical hidden Markov models},
  author={Peng, Ya-Ti and Lin, Ching-Yung and Sun, Ming-Ting and Tsai, Kun-Cheng},
  booktitle={Proc. IEEE International conference on multimedia and expo},
  pages={1218--1221},
  year={2009},
  organization={IEEE}
}

@article{kiranyaz2006generic,
  title={A generic audio classification and segmentation approach for multimedia indexing and retrieval},
  author={Kiranyaz, Serkan and Qureshi, Ahmad Farooq and Gabbouj, Moncef},
  journal={IEEE Transactions on Audio, Speech, and Language Processing},
  volume={14},
  number={3},
  pages={1062--1081},
  year={2006},
  publisher={IEEE}
}

@inproceedings{xu2018large,
  title={Large-scale weakly supervised audio classification using gated convolutional neural network},
  author={Xu, Yong and Kong, Qiuqiang and Wang, Wenwu and Plumbley, Mark D},
  booktitle={Proc. IEEE international conference on acoustics, speech and signal processing (ICASSP)},
  pages={121--125},
  year={2018},
  organization={IEEE}
}

@article{xiao2022continual,
  title={Continual Learning For On-Device Environmental Sound Classification},
  author={Xiao, Yang and Liu, Xubo and King, James and Singh, Arshdeep and Chng, Eng Siong and Plumbley, Mark D and Wang, Wenwu},
  journal={arXiv preprint:2207.07429},
  year={2022}
}

@inproceedings{huang2022progressive,
  title={Progressive continual learning for spoken keyword spotting},
  author={Huang, Yizheng and Hou, Nana and Chen, Nancy F},
  booktitle={Proc. IEEE International Conference on Acoustics, Speech and Signal Processing (ICASSP)},
  pages={7552--7556},
  year={2022},
  organization={IEEE}
}

@article{liu2022learning,
  title={Learning the Spectrogram Temporal Resolution for Audio Classification},
  author={Liu, Haohe and Liu, Xubo and Kong, Qiuqiang and Wang, Wenwu and Plumbley, Mark D},
  journal={arXiv preprint:2210.01719},
  year={2022}
}

@article{mai2022online,
  title={Online continual learning in image classification: An empirical survey},
  author={Mai, Zheda and Li, Ruiwen and Jeong, Jihwan and Quispe, David and Kim, Hyunwoo and Sanner, Scott},
  journal={Neurocomputing},
  volume={469},
  pages={28--51},
  year={2022},
  publisher={Elsevier}
}

@inproceedings{prabhu2020gdumb,
  title={Gdumb: A simple approach that questions our progress in continual learning},
  author={Prabhu, Ameya and Torr, Philip HS and Dokania, Puneet K},
  booktitle={Proc. European Conference on Computer Vision(ECCV)},
  pages={524--540},
  year={2020},
  organization={Springer}
}

@article{awasthi2021teaching,
  title={Teaching keyword spotters to spot new keywords with limited examples},
  author={Awasthi, Abhijeet and Kilgour, Kevin and Rom, Hassan},
  journal={arXiv preprint:2106.02443},
  year={2021}
}

@article{2021Few,
   title={Few-Shot Keyword Spotting in Any Language},
   url={http://dx.doi.org/10.21437/Interspeech.2021-1966},
   DOI={10.21437/interspeech.2021-1966},
   journal={Interspeech 2021},
   publisher={ISCA},
   author={Mazumder, Mark and Banbury, Colby and Meyer, Josh and Warden, Pete and Reddi, Vijay Janapa},
   year={2021},
   month={Aug} }

@incollection{mccloskey1989catastrophic,
  title={Catastrophic interference in connectionist networks: The sequential learning problem},
  author={McCloskey, Michael and Cohen, Neal J},
  booktitle={Psychology of learning and motivation},
  volume={24},
  pages={109--165},
  year={1989},
  publisher={Elsevier}
}

@article{delange2021continual,
  title={A continual learning survey: Defying forgetting in classification tasks},
  author={Delange, Matthias and Aljundi, Rahaf and Masana, Marc and Parisot, Sarah and Jia, Xu and Leonardis, Ales and Slabaugh, Greg and Tuytelaars, Tinne},
  journal={IEEE Transactions on Pattern Analysis and Machine Intelligence},
  year={2021},
  publisher={IEEE}
}

@article{masana2020class,
  title={Class-incremental learning: survey and performance evaluation on image classification},
  author={Masana, Marc and Liu, Xialei and Twardowski, Bartlomiej and Menta, Mikel and Bagdanov, Andrew D and van de Weijer, Joost},
  journal={arXiv preprint:2010.15277},
  year={2020}
}

@inproceedings{bang2021rainbow,
  title={Rainbow memory: Continual learning with a memory of diverse samples},
  author={Bang, Jihwan and Kim, Heesu and Yoo, YoungJoon and Ha, Jung-Woo and Choi, Jonghyun},
  booktitle={Proc. IEEE/CVF International Conference on Computer Vision},
  pages={8218--8227},
  year={2021}
}

@inproceedings{gal2016dropout,
  title={Dropout as a bayesian approximation: Representing model uncertainty in deep learning},
  author={Gal, Yarin and Ghahramani, Zoubin},
  booktitle={Proc. International Conference on Machine Learning},
  pages={1050--1059},
  year={2016},
  organization={PMLR}
}

@article{park2019specaugment,
  title={Specaugment: A simple data augmentation method for automatic speech recognition},
  author={Park, Daniel S and Chan, William and Zhang, Yu and Chiu, Chung-Cheng and Zoph, Barret and Cubuk, Ekin D and Le, Quoc V},
  journal={arXiv preprint:1904.08779},
  year={2019}
}

@inproceedings{ko2015audio,
  title={Audio augmentation for speech recognition},
  author={Ko, Tom and Peddinti, Vijayaditya and Povey, Daniel and Khudanpur, Sanjeev},
  booktitle={Proc. Interspeech},
  year={2015}
}

@article{hinton2015distilling,
  title={Distilling the knowledge in a neural network},
  author={Hinton, Geoffrey and Vinyals, Oriol and Dean, Jeff and others},
  journal={arXiv preprint:1503.02531},
  volume={2},
  number={7},
  year={2015}
}

@inproceedings{wu2019bic,
  title={Large scale incremental learning},
  author={Wu, Yue and Chen, Yinpeng and Wang, Lijuan and Ye, Yuancheng and Liu, Zicheng and Guo, Yandong and Fu, Yun},
  booktitle={Proc. IEEE/CVF International Conference on Computer Vision},
  pages={374--382},
  year={2019}
}

@inproceedings{rebuffi2017icarl,
  title={icarl: Incremental classifier and representation learning},
  author={Rebuffi, Sylvestre-Alvise and Kolesnikov, Alexander and Sperl, Georg and Lampert, Christoph H},
  booktitle={Proc. IEEE/CVF International Conference on Computer Vision},
  pages={2001--2010},
  year={2017}
}

@article{warden2018speech,
  title={Speech commands: A dataset for limited-vocabulary speech recognition},
  author={Warden, Pete},
  journal={arXiv preprint:1804.03209},
  year={2018}
}

@article{choi2019temporal,
  title={Temporal convolution for real-time keyword spotting on mobile devices},
  author={Choi, Seungwoo and Seo, Seokjun and Shin, Beomjun and Byun, Hyeongmin and Kersner, Martin and Kim, Beomsu and Kim, Dongyoung and Ha, Sungjoo},
  journal={arXiv preprint:1904.03814},
  year={2019}
}

@article{hsu2018nr,
  title={Re-evaluating continual learning scenarios: A categorization and case for strong baselines},
  author={Hsu, Yen-Chang and Liu, Yen-Cheng and Ramasamy, Anita and Kira, Zsolt},
  journal={arXiv preprint:1810.12488},
  year={2018}}

@article{ewc,
  title={Overcoming catastrophic forgetting in neural networks},
  author={Kirkpatrick, James and Pascanu, Razvan and Rabinowitz, Neil and Veness, Joel and Desjardins, Guillaume and Rusu, Andrei A and Milan, Kieran and Quan, John and Ramalho, Tiago and Grabska-Barwinska, Agnieszka and others},
  journal={Proce. the national academy of sciences},
  volume={114},
  number={13},
  pages={3521--3526},
  year={2017},
  publisher={National Acad Sciences}
}

@inproceedings{rwalk,
  title={Riemannian walk for incremental learning: Understanding forgetting and intransigence},
  author={Chaudhry, Arslan and Dokania, Puneet K and Ajanthan, Thalaiyasingam and Torr, Philip HS},
  booktitle={Proc. the European Conference on Computer Vision (ECCV)},
  pages={532--547},
  year={2018}
}

@inproceedings{zenke2017pi,
  title={Continual learning through synaptic intelligence},
  author={Zenke, Friedemann and Poole, Ben and Ganguli, Surya},
  booktitle={Proc. International Conference on Machine Learning},
  pages={3987--3995},
  year={2017},
  organization={PMLR}
}

@article{lopez2017gradient,
  title={Gradient episodic memory for continual learning},
  author={Lopez-Paz, David and Ranzato, Marc'Aurelio},
  journal={Proc. Advances in neural information processing systems},
  volume={30},
  year={2017}
}

@article{zhang2017mixup,
  title={mixup: Beyond empirical risk minimization},
  author={Zhang, Hongyi and Cisse, Moustapha and Dauphin, Yann N and Lopez-Paz, David},
  journal={arXiv preprint:1710.09412},
  year={2017}
}

@inproceedings{nancy1,
  title={Exemplar-inspired strategies for low-resource spoken keyword search in Swahili},
  author={Chen, Nancy F and Van Tung, Pharri and Xu, Haihua and Xiao, Xiong and Ni, Chongjia and Chen, I-Fan and Sivadas, Sunil and Lee, Chin-Hui and Chng, Eng Siong and Ma, Bin and others},
  booktitle={Proc. IEEE International Conference on Acoustics, Speech and Signal Processing (ICASSP)},
  pages={6040--6044},
  year={2016},
  organization={IEEE}
}

@inproceedings{nancy2,
  title={Strategies for Vietnamese keyword search},
  author={Chen, Nancy F and Sivadas, Sunil and Lim, Boon Pang and Ngo, Hoang Gia and Xu, Haihua and Ma, Bin and Li, Haizhou and others},
  booktitle={Proc. IEEE International Conference on Acoustics, Speech and Signal Processing (ICASSP)},
  pages={4121--4125},
  year={2014},
  organization={IEEE}
}

@inproceedings{nancy3,
  title={Low-resource keyword search strategies for Tamil},
  author={Chen, Nancy F and Ni, Chongjia and Chen, I-Fan and Sivadas, Sunil and Xu, Haihua and Xiao, Xiong and Lau, Tze Siong and Leow, Su Jun and Lim, Boon Pang and Leung, Cheung-Chi and others},
  booktitle={Proc. IEEE International Conference on Acoustics, Speech and Signal Processing (ICASSP)},
  pages={5366--5370},
  year={2015},
  organization={IEEE}
}

@inproceedings{ardila2020common,
  title={Common Voice: A Massively-Multilingual Speech Corpus},
  author={Ardila, Rosana and Branson, Megan and Davis, Kelly and Kohler, Michael and Meyer, Josh and Henretty, Michael and Morais, Reuben and Saunders, Lindsay and Tyers, Francis and Weber, Gregor},
  booktitle={Proc. the Language Resources and Evaluation Conference},
  pages={4218--4222},
  year={2020}
}

@inproceedings{mcauliffe2017Forced,
  title={Montreal Forced Aligner: Trainable Text-Speech Alignment Using Kaldi.},
  author={McAuliffe, Michael and Socolof, Michaela and Mihuc, Sarah and Wagner, Michael and Sonderegger, Morgan},
  booktitle={Proc. Interspeech},
  pages={498--502},
  year={2017}
}

@article{babu2021xls,
  title={XLS-R: Self-supervised Cross-lingual Speech Representation Learning at Scale},
  author={Babu, Arun and Wang, Changhan and Tjandra, Andros and Lakhotia, Kushal and Xu, Qiantong and Goyal, Naman and Singh, Kritika and von Platen, Patrick and Saraf, Yatharth and Pino, Juan and others},
  year={2021},
  booktitle={Proc. Interspeech},
}

@article{seo2021wav2kws,
  title={Wav2kws: Transfer learning from speech representations for keyword spotting},
  author={Seo, Deokjin and Oh, Heung-Seon and Jung, Yuchul},
  journal={IEEE Access},
  volume={9},
  pages={80682--80691},
  year={2021},
}

@article{vitter1985random,
  title={Random sampling with a reservoir},
  author={Vitter, Jeffrey S},
  journal={ACM Transactions on Mathematical Software (TOMS)},
  volume={11},
  number={1},
  pages={37--57},
  year={1985}}

@techreport{Kim2021b,
    Author = "Kim, Byeonggeun and Yang, Seunghan and Kim, Jangho and Chang, Simyung",
    title = "{QTI} Submission to {DCASE} 2021: Residual Normalization for Device-Imbalanced Acoustic Scene Classification with Efficient Design",
    institution = "DCASE2021 Challenge",
    year = "2021",
    month = "June"
}

@article{kim2021broadcasted,
  title={Broadcasted residual learning for efficient keyword spotting},
  author={Kim, Byeonggeun and Chang, Simyung and Lee, Jinkyu and Sung, Dooyong},
  journal={arXiv preprint:2106.04140},
  year={2021}
}

@article{tau2019,
  author       = {Heittola, Toni and
                  Mesaros, Annamaria and
                  Virtanen, Tuomas},
  title        = {{TAU Urban Acoustic Scenes 2019, Development 
                   dataset}},
  month        = mar,
  year         = 2019,
  publisher    = {Zenodo},
  doi          = {10.5281/zenodo.2589280},
  url          = {https://doi.org/10.5281/zenodo.2589280}
}

@inproceedings{piczak2015esc,
  title={ESC: Dataset for environmental sound classification},
  author={Piczak, Karol J},
  booktitle={Proc. the ACM international conference on Multimedia},
  pages={1015--1018},
  year={2015}
}

@inproceedings{wang2019continual,
  title={Continual learning of new sound classes using generative replay},
  author={Wang, Zhepei and Subakan, Cem and Tzinis, Efthymios and Smaragdis, Paris and Charlin, Laurent},
  booktitle={Proc. IEEE Workshop on Applications of Signal Processing to Audio and Acoustics (WASPAA)},
  pages={308--312},
  year={2019}
}

@article{singh2022passive,
  title={A Passive Similarity based CNN Filter Pruning for Efficient Acoustic Scene Classification},
  author={Singh, Arshdeep and Plumbley, Mark D},
  journal={arXiv preprint:2203.15751},
  year={2022}
}

@inproceedings{awasthi2019continual,
  title={Continual learning with neural networks: A review},
  author={Awasthi, Abhijeet and Sarawagi, Sunita},
  booktitle={Proc. the ACM India Joint International Conference on Data Science and Management of Data},
  pages={362--365},
  year={2019}
}

@inproceedings{zenke2017continual,
  title={Continual learning through synaptic intelligence},
  author={Zenke, Friedemann and Poole, Ben and Ganguli, Surya},
  booktitle={Proc. International Conference on Machine Learning},
  pages={3987--3995},
  year={2017},
}

@article{xiao2022rainbow,
  title={Rainbow Keywords: Efficient Incremental Learning for Online Spoken Keyword Spotting},
  author={Xiao, Yang and Hou, Nana and Chng, Eng Siong},
  journal={arXiv preprint:2203.16361},
  year={2022}
}

@article{audiomentation,
  author       = {Iver Jordal and
                  Shahul ES and
                  Hervé BREDIN and
                  Kento Nishi and
                  Francis Lata and
                  Harry Coultas Blum and
                  Pariente Manuel and
                  akash raj and
                  Keunwoo Choi and
                  FrenchKrab and
                  Piotr Żelasko and
                  amiasato and
                  Moreno La Quatra and
                  Emmanuel Schmidbauer},
  title        = {asteroid-team/torch-audiomentations: v0.11.0},
  month        = jun,
  year         = 2022,
  publisher    = {Zenodo},
  version      = {v0.11.0},
  doi          = {10.5281/zenodo.6778064},
  url          = {https://doi.org/10.5281/zenodo.6778064}
}

@inproceedings{kamath2002multi,
  title={A multi-band spectral subtraction method for enhancing speech corrupted by colored noise},
  author={Kamath, Sunil and Loizou, Philipos},
  booktitle={Proc. IEEE International Conference on Acoustics, Speech, and Signal Processing (ICASSP)},
year={2002}
}

@article{kloek1978bayesian,
  title={Bayesian estimates of equation system parameters: An application of integration by Monte Carlo},
  author={Kloek, Tuen and Van Dijk, Herman K},
  journal={Econometrica: Journal of the Econometric Society},
  pages={1--19},
  year={1978},
  publisher={JSTOR}
}

@article{biesialska2020continual,
  title={Continual lifelong learning in natural language processing: A survey},
  author={Biesialska, Magdalena and Biesialska, Katarzyna and Costa-Jussa, Marta R},
  journal={arXiv preprint:2012.09823},
  year={2020}
}

@article{lesort2020continual,
  title={Continual learning for robotics: Definition, framework, learning strategies, opportunities and challenges},
  author={Lesort, Timoth{\'e}e and Lomonaco, Vincenzo and Stoian, Andrei and Maltoni, Davide and Filliat, David and D{\'\i}az-Rodr{\'\i}guez, Natalia},
  journal={Information fusion},
  volume={58},
  pages={52--68},
  year={2020},
  publisher={Elsevier}
}

@article{darji2017audio,
  title={Audio signal processing: A review of audio signal classification features},
  author={Darji, Mittal C},
  journal={International Journal of Scientific Research in Computer Science, Engineering and Information Technology},
  volume={2},
  number={3},
  pages={227--230},
  year={2017}
}

@inproceedings{zhuang2016unrestricted,
  title={Unrestricted Vocabulary Keyword Spotting Using LSTM-CTC.},
  author={Zhuang, Yimeng and Chang, Xuankai and Qian, Yanmin and Yu, Kai},
  booktitle={Proc. Interspeech},
  pages={938--942},
  year={2016}
}

@article{kingma2014adam,
  title={Adam: A method for stochastic optimization},
  author={Kingma, Diederik P and Ba, Jimmy},
  journal={arXiv preprint:1412.6980},
  year={2014}
}

@inproceedings{chen2014small,
  title={Small-footprint keyword spotting using deep neural networks},
  author={Chen, Guoguo and Parada, Carolina and Heigold, Georg},
  booktitle={Proc. IEEE International Conference on Acoustics, Speech and Signal Processing (ICASSP)},
  pages={4087--4091},
  year={2014},
  organization={IEEE}
}

@inproceedings{alvarez2019end,
  title={End-to-end streaming keyword spotting},
  author={Alvarez, Raziel and Park, Hyun-Jin},
  booktitle={Proc. IEEE International Conference on Acoustics, Speech and Signal Processing (ICASSP)},
  pages={6336--6340},
  year={2019},
  organization={IEEE}
}

@article{nakkiran2015compressing,
  title={Compressing deep neural networks using a rank-constrained topology},
  author={Nakkiran, Preetum and Alvarez, Raziel and Prabhavalkar, Rohit and Parada, Carolina},
  year={2015}
}

@inproceedings{pedroni2018small,
  title={Small-footprint spiking neural networks for power-efficient keyword spotting},
  author={Pedroni, Bruno U and Sheik, Sadique and Mostafa, Hesham and Paul, Somnath and Augustine, Charles and Cauwenberghs, Gert},
  booktitle={Proc. IEEE Biomedical Circuits and Systems Conference (BioCAS)},
  pages={1--4},
  year={2018},
  organization={IEEE}
}

@article{sainath2015convolutional,
  title={Convolutional neural networks for small-footprint keyword spotting},
  author={Sainath, Tara and Parada, Carolina},
  year={2015}
}

@inproceedings{tang2018deep,
  title={Deep residual learning for small-footprint keyword spotting},
  author={Tang, Raphael and Lin, Jimmy},
  booktitle={Proc. IEEE International Conference on Acoustics, Speech and Signal Processing (ICASSP)},
  pages={5484--5488},
  year={2018},
  organization={IEEE}
}

@inproceedings{ibrahim2019keyword,
  title={Keyword spotting using time-domain features in a temporal convolutional network},
  author={Ibrahim, Emad A and Huisken, Jos and Fatemi, Hamed and de Gyvez, Jose Pineda},
  booktitle={Proc. Euromicro Conference on Digital System Design (DSD)},
  pages={313--319},
  year={2019},
  organization={IEEE}
}

@article{xu2020depthwise,
  title={Depthwise Separable Convolutional ResNet with Squeeze-and-Excitation Blocks for Small-Footprint Keyword Spotting},
  author={Xu, Menglong and Zhang, Xiao-Lei},
  journal={Proc. Interspeech},
  pages={2547--2551},
  year={2020}
}

@inproceedings{coucke2019efficient,
  title={Efficient keyword spotting using dilated convolutions and gating},
  author={Coucke, Alice and Chlieh, Mohammed and Gisselbrecht, Thibault and Leroy, David and Poumeyrol, Mathieu and Lavril, Thibaut},
  booktitle={Proc. IEEE International Conference on Acoustics, Speech and Signal Processing (ICASSP)},
  pages={6351--6355},
  year={2019},
  organization={IEEE}
}

@article{howard2017mobilenets,
  title={Mobilenets: Efficient convolutional neural networks for mobile vision applications},
  author={Howard, Andrew G and Zhu, Menglong and Chen, Bo and Kalenichenko, Dmitry and Wang, Weijun and Weyand, Tobias and Andreetto, Marco and Adam, Hartwig},
  journal={arXiv preprint:1704.04861},
  year={2017}
}

@inproceedings{mittermaier2020small,
  title={Small-footprint keyword spotting on raw audio data with sinc-convolutions},
  author={Mittermaier, Simon and K{\"u}rzinger, Ludwig and Waschneck, Bernd and Rigoll, Gerhard},
  booktitle={Proc. IEEE International Conference on Acoustics, Speech and Signal Processing (ICASSP)},
  pages={7454--7458},
  year={2020},
  organization={IEEE}
}

@inproceedings{kumar2018convolutional,
  title={On Convolutional LSTM Modeling for Joint Wake-Word Detection and Text Dependent Speaker Verification.},
  author={Kumar, Rajath and Yeruva, Vaishnavi and Ganapathy, Sriram},
  booktitle={Proc. Interspeech},
  pages={1121--1125},
  year={2018}
}

@inproceedings{sundar2015keyword,
  title={Keyword spotting in multi-player voice driven games for children},
  author={Sundar, Harshavardhan and Lehman, Jill Fain and Singh, Rita},
  booktitle={Proc. Interspeech},
  year={2015}
}

@article{arik2017convolutional,
  title={Convolutional recurrent neural networks for small-footprint keyword spotting},
  author={Arik, Sercan O and Kliegl, Markus and Child, Rewon and Hestness, Joel and Gibiansky, Andrew and Fougner, Chris and Prenger, Ryan and Coates, Adam},
  journal={arXiv preprint:1703.05390},
  year={2017}
}

@article{rybakov2020streaming,
  title={Streaming keyword spotting on mobile devices},
  author={Rybakov, Oleg and Kononenko, Natasha and Subrahmanya, Niranjan and Visontai, Mirk{\'o} and Laurenzo, Stella},
  journal={arXiv preprint:2005.06720},
  year={2020}
}

@article{madani2020progen,
  title={Progen: Language modeling for protein generation},
  author={Madani, Ali and McCann, Bryan and Naik, Nikhil and Keskar, Nitish Shirish and Anand, Namrata and Eguchi, Raphael R and Huang, Po-Ssu and Socher, Richard},
  journal={arXiv preprint:2004.03497},
  year={2020}
}

@article{brown2020language,
  title={Language models are few-shot learners},
  author={Brown, Tom and Mann, Benjamin and Ryder, Nick and Subbiah, Melanie and Kaplan, Jared D and Dhariwal, Prafulla and Neelakantan, Arvind and Shyam, Pranav and Sastry, Girish and Askell, Amanda and others},
  journal={Proc. Advances in neural information processing systems},
  volume={33},
  pages={1877--1901},
  year={2020}
}

@article{devlin2018bert,
  title={Bert: Pre-training of deep bidirectional transformers for language understanding},
  author={Devlin, Jacob and Chang, Ming-Wei and Lee, Kenton and Toutanova, Kristina},
  journal={arXiv preprint:1810.04805},
  year={2018}
}

@article{huang2018music,
  title={Music transformer},
  author={Huang, Cheng-Zhi Anna and Vaswani, Ashish and Uszkoreit, Jakob and Shazeer, Noam and Simon, Ian and Hawthorne, Curtis and Dai, Andrew M and Hoffman, Matthew D and Dinculescu, Monica and Eck, Douglas},
  journal={arXiv preprint:1809.04281},
  year={2018}
}

@inproceedings{sun2019videobert,
  title={Videobert: A joint model for video and language representation learning},
  author={Sun, Chen and Myers, Austin and Vondrick, Carl and Murphy, Kevin and Schmid, Cordelia},
  booktitle={Proc. IEEE/CVF International Conference on Computer Vision},
  pages={7464--7473},
  year={2019}
}

@inproceedings{girdhar2019video,
  title={Video action transformer network},
  author={Girdhar, Rohit and Carreira, Joao and Doersch, Carl and Zisserman, Andrew},
  booktitle={Proc. IEEE/CVF International Conference on Computer Vision},
  pages={244--253},
  year={2019}
}

@article{dosovitskiy2020image,
  title={An image is worth 16x16 words: Transformers for image recognition at scale},
  author={Dosovitskiy, Alexey and Beyer, Lucas and Kolesnikov, Alexander and Weissenborn, Dirk and Zhai, Xiaohua and Unterthiner, Thomas and Dehghani, Mostafa and Minderer, Matthias and Heigold, Georg and Gelly, Sylvain and others},
  journal={arXiv preprint:2010.11929},
  year={2020}
}

@inproceedings{liu2021swin,
  title={Swin transformer: Hierarchical vision transformer using shifted windows},
  author={Liu, Ze and Lin, Yutong and Cao, Yue and Hu, Han and Wei, Yixuan and Zhang, Zheng and Lin, Stephen and Guo, Baining},
  booktitle={Proc. IEEE/CVF International Conference on Computer Vision},
  pages={10012--10022},
  year={2021}
}

@article{berg2021keyword,
  title={Keyword transformer: A self-attention model for keyword spotting},
  author={Berg, Axel and O'Connor, Mark and Cruz, Miguel Tairum},
  journal={arXiv preprint:2104.00769},
  year={2021}
}

@article{gong2021ast,
  title={Ast: Audio spectrogram transformer},
  author={Gong, Yuan and Chung, Yu-An and Glass, James},
  journal={arXiv preprint:2104.01778},
  year={2021}
}

@article{kao2022efficiency,
  title={On the Efficiency of Integrating Self-supervised Learning and Meta-learning for User-defined Few-shot Keyword Spotting},
  author={Kao, Wei-Tsung and Wu, Yuen-Kwei and Chen, Chia Ping and Chen, Zhi-Sheng and Tsai, Yu-Pao and Lee, Hung-Yi},
  journal={arXiv preprint:2204.00352},
  year={2022}
}

@article{hsu2021hubert,
  title={Hubert: Self-supervised speech representation learning by masked prediction of hidden units},
  author={Hsu, Wei-Ning and Bolte, Benjamin and Tsai, Yao-Hung Hubert and Lakhotia, Kushal and Salakhutdinov, Ruslan and Mohamed, Abdelrahman},
  journal={IEEE/ACM Transactions on Audio, Speech, and Language Processing},
  volume={29},
  pages={3451--3460},
  year={2021},
  publisher={IEEE}
}

@article{baevski2020wav2vec,
  title={wav2vec 2.0: A framework for self-supervised learning of speech representations},
  author={Baevski, Alexei and Zhou, Yuhao and Mohamed, Abdelrahman and Auli, Michael},
  journal={Advances in Neural Information Processing Systems},
  volume={33},
  pages={12449--12460},
  year={2020}
}

@article{mohamed2022self,
  title={Self-Supervised Speech Representation Learning: A Review},
  author={Mohamed, Abdelrahman and Lee, Hung-yi and Borgholt, Lasse and Havtorn, Jakob D and Edin, Joakim and Igel, Christian and Kirchhoff, Katrin and Li, Shang-Wen and Livescu, Karen and Maal{\o}e, Lars and others},
  journal={arXiv preprint:2205.10643},
  year={2022}
}

@article{zhou2022audio,
  title={Audio-Visual Wake Word Spotting in MISP2021 Challenge: Dataset Release and Deep Analysis},
  author={Zhou, Hengshun and Du, Jun and Zou, Gongzhen and Nian, Zhaoxu and Lee, Chin-Hui and Marco, Sabato},
  year={2022}
}

@inproceedings{piczak2015environmental,
  title={Environmental sound classification with convolutional neural networks},
  author={Piczak, Karol J},
  booktitle={Proc. IEEE 25\textsuperscript{th} International Workshop on Machine Learning for Signal Processing (MLSP)},
  pages={1--6},
  year={2015}
}

@inproceedings{choi2022temporal,
  title={Temporal Knowledge Distillation for On-device Audio Classification},
  author={Choi, Kwanghee and Kersner, Martin and Morton, Jacob and Chang, Buru},
  booktitle={Proc. IEEE International Conference on Acoustics, Speech and Signal Processing (ICASSP)},
  pages={486--490},
  year={2022}
}

@article{martin2022low,
  title={{Low-complexity acoustic scene classification in DCASE 2022 Challenge}},
  author={Mart{\'\i}n-Morat{\'o}, Irene and Paissan, Francesco and Ancilotto, Alberto and Heittola, Toni and Mesaros, Annamaria and Farella, Elisabetta and Brutti, Alessio and Virtanen, Tuomas},
  journal={arXiv preprint:2206.03835},
  year={2022}
}

@techreport{Singh2022,
    Author = "Singh, Arshdeep and A King, James and Liu, Xubo and Wang, Wenwu and D. Plumbley, Mark",
    title = "Low-Complexity {CNNs} for Acoustic Scene Classification",
    institution = "DCASE2022 Challenge",
    year = "2022",
    month = "June"
}

@techreport{Liu2022a,
    Author = "Liu, Haohe and Liu, Xubo and Mei, Xinhao and Kong, Qiuqiang and Wang, Wenwu and Plumbley, Mark D",
    title = "SURREY SYSTEM FOR \text{DCASE} 2022 TASK 5 : FEW-SHOT BIOACOUSTIC EVENT DETECTION WITH SEGMENT-LEVEL METRIC LEARNING Technical Report",
    institution = "DCASE2022 Challenge",
    year = "2022",
    month = "June"
}

@inproceedings{gygi2007environmental,
  title={Environmental sound research as it stands today},
  author={Gygi, Brian and Shafiro, Valeriy},
  booktitle={Proc. Meetings on Acoustics 153ASA},
  volume={1},
  number={1},
  pages={050002},
  year={2007},
  organization={Acoustical Society of America}
}

@article{gaver1993world,
  title={What in the world do we hear?: An ecological approach to auditory event perception},
  author={Gaver, William W},
  journal={Ecological psychology},
  volume={5},
  number={1},
  pages={1--29},
  year={1993},
  publisher={Taylor \& Francis}
}

@article{mesaros2017detection,
  title={Detection and classification of acoustic scenes and events: Outcome of the DCASE 2016 challenge},
  author={Mesaros, Annamaria and Heittola, Toni and Benetos, Emmanouil and Foster, Peter and Lagrange, Mathieu and Virtanen, Tuomas and Plumbley, Mark D},
  journal={IEEE/ACM Transactions on Audio, Speech, and Language Processing},
  volume={26},
  number={2},
  pages={379--393},
  year={2017},
  publisher={IEEE}
}

@article{mesaros2019sound,
  title={Sound event detection in the DCASE 2017 challenge},
  author={Mesaros, Annamaria and Diment, Aleksandr and Elizalde, Benjamin and Heittola, Toni and Vincent, Emmanuel and Raj, Bhiksha and Virtanen, Tuomas},
  journal={IEEE/ACM Transactions on Audio, Speech, and Language Processing},
  volume={27},
  number={6},
  pages={992--1006},
  year={2019},
  publisher={IEEE}
}

@article{cakir2017convolutional,
  title={Convolutional recurrent neural networks for polyphonic sound event detection},
  author={Cak{\i}r, Emre and Parascandolo, Giambattista and Heittola, Toni and Huttunen, Heikki and Virtanen, Tuomas},
  journal={IEEE/ACM Transactions on Audio, Speech, and Language Processing},
  volume={25},
  number={6},
  pages={1291--1303},
  year={2017},
  publisher={IEEE}
}

@inproceedings{ren2019attention,
  title={Attention-based atrous convolutional neural networks: Visualisation and understanding perspectives of acoustic scenes},
  author={Ren, Zhao and Kong, Qiuqiang and Han, Jing and Plumbley, Mark D and Schuller, Bj{\"o}rn W},
  booktitle={Proc. IEEE International Conference on Acoustics, Speech and Signal Processing (ICASSP)},
  pages={56--60},
  year={2019},
  organization={IEEE}
}

@inproceedings{koutini2019cp,
  title={CP-JKU submissions to DCASE’19: Acoustic scene classification and audio tagging with receptive-field-regularized CNNs},
  author={Koutini, Khaled and Eghbal-zadeh, Hamid and Widmer, Gerhard and Kepler, J},
  booktitle={Proc. the Detection and Classification of Acoustic Scenes and Events Workshop (DCASE)},
  pages={25--26},
  year={2019}
}

@inproceedings{yang2018acoustic,
  title={Acoustic scene classification using multi-scale features.},
  author={Yang, Liping and Chen, Xinxing and Tao, Lianjie},
  booktitle={Proc. the Detection and Classification of Acoustic Scenes and Events Workshop (DCASE)},
  pages={29--33},
  year={2018}
}

@article{cho2019acoustic,
  title={Acoustic scene classification based on a large-margin factorized CNN},
  author={Cho, Janghoon and Yun, Sungrack and Park, Hyoungwoo and Eum, Jungyun and Hwang, Kyuwoong},
  journal={arXiv preprint arXiv:1910.06784},
  year={2019}
}

@inproceedings{wu2017asymmetrie,
  title={Asymmetrie Kernel Convolutional Neural Network for acoustic scenes classification},
  author={Wu, Yu-Chi and Chang, Pao-Chi and Wang, Chien-Yao and Wang, Jia-Ching},
  booktitle={Proc. IEEE International Symposium on Consumer Electronics (ISCE)},
  pages={11--12},
  year={2017},
  organization={IEEE}
}

@inproceedings{basbug2019acoustic,
  title={Acoustic scene classification using spatial pyramid pooling with convolutional neural networks},
  author={Basbug, Ahmet Melih and Sert, Mustafa},
  booktitle={Proc. IEEE International Conference on Semantic Computing (ICSC)},
  pages={128--131},
  year={2019},
  organization={IEEE}
}

@inproceedings{Marchi2016,
    Author = "Marchi, Erik and Tonelli, Dario and Xu, Xinzhou and Ringeval, Fabien and Deng, Jun and Squartini, Stefano and Schuller, Bjoern",
    title = "Pairwise Decomposition with Deep Neural Networks and Multiscale Kernel Subspace Learning for Acoustic Scene Classification",
    booktitle = "the Detection and Classification of Acoustic Scenes and Events Workshop (DCASE)",
    year = "2016",
    month = "September",
    pages = "65--69",
}

@inproceedings{Bisot2017NonnegativeFL,
  title={Nonnegative Feature Learning Methods for Acoustic Scene Classification},
  author={Victor Bisot and Romain Serizel and Slim Essid and Ga{\"e}l Richard},
  year={2017}
}

@INPROCEEDINGS{8282314,  author={Takahashi, Gen and Yamada, Takeshi and Ono, Nobutaka and Makino, Shoji},  booktitle={Proc. Asia-Pacific Signal and Information Processing Association Annual Summit and Conference (APSIPA ASC)},   title={Performance evaluation of acoustic scene classification using DNN-GMM and frame-concatenated acoustic features},   year={2017},  volume={},  number={},  pages={1739-1743},  doi={10.1109/APSIPA.2017.8282314}}

@article{kong2020panns,
  title={Panns: Large-scale pretrained audio neural networks for audio pattern recognition},
  author={Kong, Qiuqiang and Cao, Yin and Iqbal, Turab and Wang, Yuxuan and Wang, Wenwu and Plumbley, Mark D},
  journal={IEEE/ACM Transactions on Audio, Speech, and Language Processing},
  volume={28},
  pages={2880--2894},
  year={2020},
  publisher={IEEE}
}

@article{gong2021psla,
  title={Psla: Improving audio tagging with pretraining, sampling, labeling, and aggregation},
  author={Gong, Yuan and Chung, Yu-An and Glass, James},
  journal={IEEE/ACM Transactions on Audio, Speech, and Language Processing},
  volume={29},
  pages={3292--3306},
  year={2021},
  publisher={IEEE}
}

@article{li2018attention,
  title={An attention pooling based representation learning method for speech emotion recognition},
  author={Li, Pengcheng and Song, Yan and McLoughlin, Ian Vince and Guo, Wu and Dai, Li-Rong},
  year={2018},
  publisher={International Speech Communication Association}
}

@inproceedings{gong2022ssast,
  title={Ssast: Self-supervised audio spectrogram transformer},
  author={Gong, Yuan and Lai, Cheng-I and Chung, Yu-An and Glass, James},
  booktitle={Proc. the AAAI Conference on Artificial Intelligence},
  volume={36},
  number={10},
  pages={10699--10709},
  year={2022}
}

@inproceedings{gemmeke2017audio,
  title={Audio set: An ontology and human-labeled dataset for audio events},
  author={Gemmeke, Jort F and Ellis, Daniel PW and Freedman, Dylan and Jansen, Aren and Lawrence, Wade and Moore, R Channing and Plakal, Manoj and Ritter, Marvin},
  booktitle={Proc. IEEE international conference on acoustics, speech and signal processing (ICASSP)},
  pages={776--780},
  year={2017},
  organization={IEEE}
}

@article{li2017learning,
  title={Learning without forgetting},
  author={Li, Zhizhong and Hoiem, Derek},
  journal={IEEE transactions on pattern analysis and machine intelligence},
  volume={40},
  number={12},
  pages={2935--2947},
  year={2017},
  publisher={IEEE}
}
